\documentclass[12pt,preprint]{aastex} 
 
\shorttitle{Dwarf galaxies in Coma cluster} 
\shortauthors{Aguerri et al.} 
 
\begin{document} 
 
 
\title{Structural Parameters of Dwarf Galaxies in the Coma Cluster:
  On the Origin of dS0 Galaxies.} 
 
 
\author{J. A. L. Aguerri} 
\affil{Instituto de Astrof\'{\i}sica de Canarias, C/ V\'{\i}a L\'actea s/n,
 38205 La Laguna, Spain} 
 
\author{J. Iglesias-P\'aramo} 
\affil{Laboratoire d'Astrophysique de Marseille, BP8, Traverse du Siphon, 13376
 Marseille, France} 
\affil{Instituto de Astrof\'{\i}sica de Andaluc\'{\i}a, Apdo. 3004, 18080
 Granada, Spain  } 

\author{J. M. V\'{\i}lchez} 
\affil{Instituto de Astrof\'{\i}sica de Andaluc\'{\i}a, Apdo. 3004, 18080
 Granada, Spain  } 
 
\and  
 
\author{C. Mu\~noz-Tu\~n\'on \& R. S\'anchez-Janssen} 
\affil{Instituto de Astrof\'{\i}sica de Canarias, V\'{\i}a L\'actea s/n, 38205
 La Laguna. Spain.} 
 
 
 

 
\begin{abstract} 
 In this paper we analyze the structural parameters of the dwarf
 galaxies in the Coma cluster with $-18 \le M_{B} \le -16$ and
 classify them 
 in two types: those with surface brightness profiles well
fitted by a single Sersic law were called dwarf ellipticals (dEs),
and those fitted with Sersic-plus-exponential profiles were
classified as dwarf lenticulars (dS0s). The comparison of the structural
parameters of the dwarf galaxies in the Coma and Virgo clusters shows that they are
analogous. Photometrically, the dE and dS0 galaxies in Coma are equivalent, having
 similar colors and global scales. However, the scale of the
 innermost parts (bulges) of dS0 galaxies is similar to the bulges of
 late-type spiral galaxies. In contrast, dEs have larger scales than the bulges
 of bright galaxies. This may indicate that dS0 and dE galaxies have different
origins. While dE galaxies can come from dwarf irregulars
(dIs) or from similar processes as bright Es, the origin of dS0
galaxies can be harassed bright late-type spiral galaxies.

\end{abstract} 
 
 
 
 
\section{Introduction}
 
Dwarf galaxies are the most numerous type of galaxies in the Universe. The term
 dwarf is applied to galaxies with low central surface brightness and
 with luminosities below $M_{B}=-18$. According to the hierarchical theory of
 galaxy evolution, dwarfs are the building blocks of the brightest galaxies. We can
 distinguish two main classes of dwarf galaxies: dwarf ellipticals (dEs) and dwarf
 Irregulars (dIs). These low mass systems have similar stellar
 distributions, both in terms of functional form and in terms of
 typical central surface brightness and scale lengths (Lin \& Faber
 1983; van Zee et al. 2004a). In general, their surface brightness profiles can be
 well fitted by Sersic (1968) profiles, with shape parameter $n
 \approx$ 1--2 (Barazza et al 2003; Graham \& Guzm\'an 2003).
 However, some of the dE galaxies show a nucleus in the center of the object that is well fitted by a Gaussian
 point-like source (Graham \& Guzm\'an 2003). Both types also follow the
 same luminosity--metallicity relation (Skillman et al. 1989; Richer \&
 McCall 1995). Nevertheless, they differ in  gas content and the age of their
 stellar populations,   dIs being
 gas-rich galaxies with active star formation (Patterson \& Thuan
 1996; van Zee et al. 2001)
 in contrast with dEs, which are gas-poor objects with an old and
 evolved stellar population. These differences in  gas content and
 the star formation are also presented in the typical broad-band colors of these
 objects. Thus, dI galaxies typically have $B-R \approx 1$ and dEs 
 $B-R\approx 1.5$ (Trentham 1998). 

Sandage \& Binggeli (1984) observed
 that the surface brightness profiles of some of the dE galaxies in
 Virgo cluster could not be fitted with a
 single component. Binggeli \& Cameron (1993) fitted an exponential
 profile to the outermost regions of the surface brightness of these
 galaxies, but this clearly cannot fit their surface brightness
 profiles at all radii, which indicates that these objects have at
 least two photometric components. They called 
 these types of objects dwarf lenticular (dS0) galaxies. These galaxies
  are also gas-poor systems and have an evolved and old stellar
 population, with  colors typical of dE galaxies, but they present
 higher flattening (Binggeli \& Popescu 1995).  It has recently been
 discovered that some of the dS0 galaxies in Virgo cluster have
 a disk-like structure, due to the presence of bars or/and spiral
 patterns  (Jerjen et al. 2000; Barazza el al. 2002).

In the standard picture, dwarf galaxies are formed from the gravitational
 collapse of primordial density fluctuations. Once the first stars are formed,
 mechanisms of energy feedback into the interstellar medium are proposed in
 order to regulate the subsequent star formation or even  change in the
 structure of the galaxy (Dekel \& Silk 1986). It has been observed by
 several authors in the past that the structural
 parameters of dEs and bright E galaxies follow a continuous and
 global relation. This argument has been used in order to infer a
 common formation process for dE and bright E galaxies,  dEs being the
 genuine low luminosity extension of giant E galaxies (Binggeli 1985;
 Jerjen \& Binggeli 1997; Graham \& Guzm\'an 2003). In contrast, other authors,  on the basis of a
 similar stellar distribution of dE and
 dI galaxies, have suggested the presence of linked evolutionary scenarios
 between both types of galaxies (Lin \& Faber 1983; Kormendy 1985;
 Caldwell \& Bothun 1987; Binggeli \& Cameron 1991). This last
 framework has been  reinforced by recent kinematic studies
 of dE galaxies in the Virgo cluster (Pedraz et
 al.\.\ 2002; Geha et al.\ 2002; Geha et al.\ 2003; van Zee et al.\ 2004a). It was found that most of the
 dE galaxies are rotationally supported in a similar to  dIs. This indicates that dE
 galaxies can be evolved dIs that have lost their gas
 content. The important question is then how or why dI galaxies stopped
 their star formation and lost their gas content. Several physical
 mechanisms have been proposed for sweeping the gas and stopping the star formation
 in a dI galaxy to produce a dE. These can be classified into internal
 (kinetic energy from supernova explosions; Dekel \& Silk 1986; De
 Young \& Gallagher 1990) or external processes (related with
 environmental processes in high density environments like galaxy
 clusters; Lin \& Faber 1983; van Zee et al.\ 2004a,b). The former
 explanation, however, deserves  more detailed modeling. As shown in
 Silich \& Tenorio-Tagle (2001), more parameters, such as the structure
 of the host galaxy, have to be taken into account in order discover the likely
  fate of the material processed in the starbusts.  

Galaxies in high density environments, like galaxy clusters, evolve through
many different physical processes; e.g., harassment (Moore et
al. 1996), ram-pressure stripping (Gunn \& Gott 1972; Quilis et
al. 2000), tidal effects and mergers (Toomre \& Toomre 1972; Bekki et
al. 2001; Aguerri et al. 2001) or starvation (Bekki et al.\ 2002). In
Aguerri et al. (2004) it was shown that bright spiral galaxies in the
Coma
cluster present smaller disk scale-lengths than equivalent field
galaxies. This was explained in terms of  the harassment suffered by the
galaxies in Coma while they are falling into the cluster. But  the above cited physical
mechanisms  not only play an important role in the evolution of bright
galaxies, they can even strongly affect the evolution of dwarf
galaxies. They are mostly located in galaxy clusters, being more
common in lower density environments (Phillips et al. 1998, Pracy et
al. 2004; S\'anchez-Janssen et al. 2005), which indicates that there
is a link between the dwarf population and the cluster environment.  Van Zee et al. (2004a) argued that the evolution from dIs to
dE galaxies was mostly driven by the gas stripping mechanism. Thus, dI
galaxies falling into the cluster potential lose their gas content by
 shock with the hot intracluster medium, evolving into dE
galaxies. This mechanism can explain why dIs and dEs have similar
scales, as well as the rotation features recently observed in some dE galaxies. This
scenario could also explain the evolution of dIs into dE galaxies without
rotation features. As demonstrated by Mayer et al.\ (2001a,b),
repeated tidal interactions suffered by a dwarf satellite galaxy can
remove the kinetic signature and transform a dI galaxy into a dE.
 But  more than one single mechanism can drive this
evolution, it being difficult to distinguish which of them is the
dominant one. Recently, the development of numerical
simulations have provided a way of following the evolution of galaxies in
cluster-like potentials (Moore et al.\ 1996, 1998, 1999, 2000). The main conclusion of these studies is
 that the morphological evolution of galaxies in clusters is driven, on short
 timescales, by the interactions between galaxies and by the gravitational
 cluster potential. The cumulative effect of these encounters can cause a
 dramatic morphological transition between late type spiral galaxies to dwarfs
 (Moore et al.\ 1996). This has been recently confirmed by Mastropietro
 et al. (2004), who  have studied the evolution of a late-type galaxy
 in a galaxy cluster simulation and found that the galaxies
 undergo significant transformations,  moving through the Hubble
 sequence from late type to dwarf galaxies. So, this could be another mechanism for the formation
 of dwarf galaxies in clusters. In this  paper we  analyze the structural parameters of a
sample of dwarf galaxies in the Coma cluster and   study their
relation with the different evolutionary scenarios described.

The population of
 dwarf galaxies in Coma  has been extensively studied  in the
 literature, but most of these researches are related with properties
 of their luminosity function (Thompson \& Gregory 1993; Trentham 1998;
 Iglesias-P\'aramo et al. 2002; Mobasher et
 al. 2003), or their color and stellar population
 analysis (Secker et al. 1997; Trentham 1998; Poggianti et al. 2001). 
 In contrast to other nearby clusters like Virgo, very little 
 is reported in the literature about the structural parameters of dwarf
 galaxies in Coma. Graham \& Guzm\'an (2003) have recently published the
 structural parameters of a sample of 16 dwarf galaxies in Coma using
 {\itshape HST} images. Previous studies of morphology in  the Coma
 cluster galaxies do not reach the dwarf luminosities (Andreon et
 al. 1996, 1997). Only Gutierrez et al. (2004) give the quantitative
 morphology of Coma galaxies down to $M_{B}=-16$, but restricted to the central
 0.25 $\deg^{2}$ region. In the present paper we present the largest
 study of structural parameters for dwarf galaxies in Coma. We have analyzed all dwarf galaxies with
 luminosities brighter than $M_{B}=-16$, located in an area of 1
 $\deg^{2}$ around the center of the cluster. 

This paper is organized as follows. In Section 2 we present the sample of dwarf
 galaxies analyzed. The fit of the surface brightness profiles of the objects is shown
 in section 3. The correlation between the structural parameters and
 the comparison with dwarf galaxies in Virgo  are shown in
 Section 4. The discussion is presented in Section 5, and conclusions
 are given in Section 6. Through this paper we will assume
 $H_{0}=75$ km s$^{-1}$ Mpc$^{-1}$.

\section{Observations and Sample Selection} 
 
The observations were taken with the Wide Field Camera (WFC) at the prime focus
 of the  2.5 m Isaac Newton Telescope (INT)  at the Roque de los Muchachos
 Observatory (La Palma), in  2000 April. The plate-scale of the CCD was 0.333
 arcsec/pixel, and the seeing of the images was about 1.5$''$. 

Although a detailed description of the observations, data
    reduction, calibration and extraction of the objects are given in
    Iglesias-P\'aramo et al. (2002, 2003), a brief summary is given
    here. Four fields of the Coma cluster were obtained in photometric
    conditions, covering and area of $\approx 1\ \deg^{2}$ in the
    North-East region of the cluster, coinciding with the
 central part of the Godwin catalog of the Coma cluster (Godwin et
 al. 1983). Three different exposures of 300 s each, slightly
    dithered, 
 to remove cosmic rays, were obtained for each field. The data
 reduction of the images was carried out using standard IRAF
 tasks, and the images were astrometrized using the United States
    Naval Observatory (USNO) catalog of stars. After the calibration
    of the images, typical errors (Poisson and zero point uncertainty)
    in the magnitude of the objects were about 0.15 mag.

We have selected as dwarf galaxies those with apparent magnitude brighter than
 $m_{r}=17$, absolute magnitude
 fainter than $M_{B}=-18$\footnote{The adopted distance modulus of the
 cluster was -34.83, this was obtained assuming a recessional velocity
 of 6925 km s$^{-1}$ obtained from NASA Extragalactic Database (NED)}, and belonging
 to the Coma cluster. The limiting magnitude $m_{r}=17$ was taken
 based on Monte Carlo simulations, in order to determine the limiting
 magnitude for a reliable measurement of the structural parameters of
 the galaxies (see Aguerri et al. 2004). We have obtained the
 recessional velocities and the $B$ band magnitudes of the objects
 from Godwin et al. \ (1983) catalog. We considered as Coma cluster
 members those galaxies with 4000 $\leq cz \leq$ 10000.
 This gives a total number of 99 objects with absolute magnitudes in
 the range $-16\le M_{B} \le -18$. All the galaxies selected in
 this study have a color $B-R>1.25$. This is the typical color shown
 by an evolved and old stellar population like that of dE galaxies (see Graham \& Guzm\'an 2003).
 
We fitted ellipses to the isophotes of all the sample, 
 using the IRAF task  ELLIPSE. This provides the radial surface
 brightness profiles of the objects, which were then fitted by 
 mathematical functions (see Section 3). In 15\% of the galaxies we
 could not fit the isophotes, because  these objects were located
 very close to a bright galaxy or field star, and no reliable
 photometry could be obtained for them.

\section{Fit of the Surface Brightness Profiles} 
 
The surface brightness profiles of the objects were fitted using a Sersic
  or Sersic+exponential profiles. The Sersic profile is given by the
 law (Sersic 1968): 
 
\begin{equation} 
\mu(r)=\mu_{\rm e} +2.5 \times b_{n}((\frac{r}{r_{\rm e}})^{1/n}-1), 
\end{equation} 
where $r_{\rm e}$ is the effective radius, which encloses half of the
total luminosity  
of the profile, $\mu_{\rm e}$ is the effective surface brightness, and $n$ is the profile 
shape parameter. The parameter 
$b_{n}$ is given by $b_{n}=0.868n-0.142$ (Caon et al. 1993).  
 
The exponential profile is given by (Freeman 1970): 
 
\begin{equation} 
\mu(r)=\mu_{0}-1.0857 \times r/h, 
\end{equation} 
where $\mu_{0}$ is the central surface brightness and $h$ is the scale length 
of the profile. 
 
The free parameters of the models were fitted using a Levenberg--Marquardt nonlinear
 fitting routine and  were determined by minimizing the $\chi^{2}$ of the fit.
 We have used the algorithm designed by Trujillo et al. (2001a). This
algorithm takes into account the seeing effects on the surface
brightness profiles (Trujillo et al. 2001b,c) and the intrinsic ellipticity of the
 galaxies. It has  been successfully applied in previous studies of the
 quantitative morphology of galaxies in nearby clusters (Trujillo et
 al. 2002; Gutierrez et al. 2004; Aguerri et al. 2004), medium
 redshift clusters (Trujillo et al. 2001a), and field galaxies (Aguerri
 \& Trujillo 2002; Trujillo \& Aguerri 2004).

Extensive
 Monte Carlo simulations were carried out in order to determine the uncertainties
in the determination of the structural parameters. We simulated
 galaxies with structural parameters similar to the observed objects. These simulations tell us
 that the structural parameters of simulated objects with two
 components can be obtained
 within errors less than 20$\%$ for those objects with magnitudes brighter
 than $m_{r}=17$. Objects modeled with only one component can be recovered
 with errors less than 20$\%$ until $m_{r}=19$, (see the simulations presented
 in Aguerri et al.\ 2004).

All the galaxies were first fitted with a single Sersic profile. If
the residuals from this fit (taking into account the photometric
errors in the surface brightness profiles) were always less than 0.15 mag
arcsec$^{-2}$, then we took such fit as a 
good one. Those galaxies with larger residuals were fitted with two
components (Sersic+exponential). Figure 1 shows, as an example, the
fits and residuals of two galaxies. It can be seen that the fits of
those objects with one component have very large residuals in the
outermost regions of their surface brightness profiles. In contrast, 
the residuals were reduced when two components were
fitted. This fitting procedure ensures a fit to
the surface brightness profiles of the galaxies using the smallest
number of components.  

The galaxies were classified according with the number of fitted
components. Those well
fitted by a single Sersic  profile, were called dE
galaxies, and those fitted with Sersic+exponential profiles were
called dS0 galaxies. The result was 55 objects classified as dEs 
 and 29 as dS0 galaxies. Tables 1 and 2 show the structural parameters
 of the dE and dS0 galaxies.  Figures 2 and 3 show the fit of the surface brightness profiles of
 the galaxies and the
 residuals. 

\subsection{Comparison with previous studies}

The goodness of the fit of the surface brightness
profiles of our galaxies can be analyzed by comparing the aperture magnitude of the
 objects obtained with SExtractor (Bertin \& Arnouts 1996), and those computed from the fitted
 models in this work. This comparison is showed in Fig. 4, where it can be
 seen that the agreement between both magnitudes is very good in all cases. 

The structural parameters of dwarf galaxies in the Coma cluster have not
been extensively studied in the literature. Actually, there are only two
previous works with which we can compare our results: Gutierrez et
al. (2004) and Graham \& Guzm\'an (2003).

Gutierrez et al. (2004) obtained the structural parameters of Coma
galaxies located in the central part (0.25 deg$^{2}$) of the
cluster down to $m_{r}=17$. They used the same code for fitting the surface brightness
profiles, but, their classification of the galaxies was slightly
different.\footnote{Gutierrez et al. (2004) fitted all the galaxies
  with two components (Sersic+exponential), and those with
a  bulge-to-disk (B/T) ratio larger than 0.6 or not well fitted with
  two components were fitted with a single Sersic profile.}  We have
53 galaxies in common with the Gutierrez et
al. (2004) sample. The same classification in both samples is achieved
in $74\%$ of the common objects. The mean errors of the structural parameters
for those galaxies are less than 20$\%$, which is the typical error
obtained from the simulations for the structural parameters of these
objects (see Aguerri et al. 2004).  For the rest, 12 were
classified as dS0s by Gutierrez et al.\ (2004), but our fits prove that
only one Sersic profile is a good fit for those objects. Two of them,
GMP 3126 and GMP 3463, have structure in the outermost part of the
surface brightness profiles, and only the inner regions of the
profiles were fitted, and two more (GMP2615, GMP 2778) have unreliable
parameters in Gutierrez et al. (2004).  If we exclude this four
galaxies, them we have the same classification in both samples for the
80$\%$ of the common galaxies. There are also two galaxies (GMP 3017 and GMP
3707) classified as dEs by
Gutierrez et al.\ and as dS0s in this study. But, for these two galaxies one
component can not fit the external parts of their surface brightness
profiles. For this reason we include a second component in the fit.

It should be noticed that the inclusion of a second
component in the fits would reduce the $\chi^{2}$ and the residuals
of the fits due to the addition of more free parameters. We have only
fitted a second component in those cases were the residuals were very
large, especially in the outermost regions of the profiles. This means
that the number of dS0 galaxies classified in this study is a lower
limit. It is possible that some of the galaxies classified as dEs have two components. 
Future kinematic studies of these objects will shed some
light on this problem.

There is  another important difference from Gutierrez et
al.\ (2004). Although we have used the same code for the decomposition
of the surface brightness profiles of the galaxies, we have not used
the same images. This means that we have not fitted the structural
parameters of the galaxies to the same surface brightness
profiles. Variations in the sky background level of about 1--2$\%$ 
can affect to the external parts of the surface brightness profiles of
the objects and modify by about 10--20$\%$ the fitted structural parameters of
the galaxies. This effect is especially important for such faint
objects  as those studied in the present paper. This is another
reason why we have not obtained exactly the same parameters for those
galaxies classified as dEs in Gutierrez et al.\ (2004) and the present
paper.   

We have five galaxies in common with Graham \& Guzm\'an (2003). GMP 2960 and GMP 3292 were classified in our and
 Graham \& Guzm\'an (2003) samples as dS0. The differences in the
 parameters for GMP 2960 are about 30$\%$. GMP 3292 has larger
 differences in the parameters of the bulge component because Graham \&
 Guzm\'an (2003) fitted a Gaussian profile in the innermost region of
 this galaxy; the disk is very similar to our fitted disk. The
 galaxies GMP 2879 and GMP 2985 were fitted with  single Sersic
 profiles in both samples, and the mean error in the structural
 parameters is less than 20$\%$. We have fitted the galaxy GMP 3486
 with a single Sersic profile, while Graham \& Guzm\'an (2003) fitted
 it 
 with a two component fit. But, our fit demonstrates that the residuals
 of the fit are good enough with only one
 component. The differences of the fitted structural
 parameters for the galaxies in common with Graham \& Guzm\'an (2003)
 could be  due to the high resolution and small background level of the {\itshape HST}
 images.

\subsection{Contamination of the sample}

Because of their low luminosity, the classification of these objects is not easy. As mentioned before, the sky background
subtraction or the presence of close companions can influence the
fit of the surface brightness profiles and the fitted structural
parameters of the objects. There is a further concern about the
    selection of the sample related on the fact that we can also have some level of
    contamination by spirals due to the uncertainty in the Coma
    cluster distance and the depth of the cluster. Spiral galaxies
    located at the back of the cluster or in the background might be
    taken as dS0s at the front of the cluster. We have visually
    checked out the images of the dS0 galaxies (see Fig. 5) in order to
    reduce the possible contamination by spirals. It can be seen that
    only one galaxy (GMP 2914) can be a spiral galaxy, instead of a
    dS0. This galaxy shows low surface brightness spiral arms and a
    strong twist of the inner and outer isophotes. The rest of the
    galaxies in the 
    sample show no signs of spiral structure.

In order to test how the results presented
here depend on the classification of the galaxies, we have selected a
subsample of dS0 galaxies.They are the 12 objects with the highest
values of 
$\chi^{2}$ for the fits of their surface brightness profiles by only
one Sersic component. These objects are classified in our sample and
Gutierrez et al (2004) sample as galaxies with two photometric
components. These will be called the ''more reliable sample'' and will be shown with larger
symbols in all plots where dS0 galaxies will be plotted. We have
    exclude from this sample the possible spiral galaxy GMP 2914.
 
\section{Correlations among the structural parameters}

\subsection{Comparison with dwarf galaxies in the Virgo cluster}

The quantitative morphology of dwarf galaxies in Virgo  has
been investigated previously by several authors in the literature. We have compared the
structural parameters of dEs in the Coma cluster with those  in the Virgo
cluster given by Durrell (1997), Barazza et al. (2003) and van Zee et
al. (2004a,b). Figure 6 shows $r_{\rm e}$, $n$ and $\mu_{o,R}$ as function of
the $B$-band absolute magnitude for dE galaxies in the Coma and Virgo clusters.  It can be
seen that Virgo and Coma dE galaxies share the same locus in the plots
for the common magnitude interval of the data: $-17\le M_{B}\le -16$. We have run 2D Kolmogorov--Smirnoff (KS2D) tests in order
to compare the distribution of our data with that in the
literature for Virgo cluster (also shown in Figure 6).  In all
cases the tests give that the distributions of points in the magnitude
interval $-17\le M_{B}\le -16$ are not statistically
different. Figure 5 also shows a trend in the structural
parameters of dEs. Thus, fainter dE galaxies have smaller $r_{\rm e}$ and
$n$, and show fainter $\mu_{o}$. The Pearson correlation coefficients
($r$) of $r_{\rm e}-M_{B}$, $n-M_{B}$ and $\mu_{o}-M_{B}$ relations are:
$-$0.30, $-$0.42, and $-$0.13, respectively. The significance of the
correlations\footnote{The significance of the correlation is the
  probability that $|r|$ should be larger than its observed value in
  the null hypothesis.} ($P$) are: 0.002, 0.001, and 0.17, respectively.

The comparison of the structural parameters of our dS0 galaxies and
those in the Virgo cluster from the literature is difficult because of the
different fitting techniques used for fitting their surface brightness profiles. Binggeli \&
Cameron (1993) fitted an exponential profile in the external parts of
the surface brightness of the dS0 galaxies in Virgo. We have compared the
scale-lengths and central surface brightness of the exponential fits
of Binggeli \& Cameron (1993) for dS0s in Virgo with the scale-lengths
and central surface brightness of our exponential component in our dS0
galaxies in the Coma cluster.  Figure 7 shows
the results.\footnote{The $B$-band central surface brightness of our exponential
fits was obtained from the $B-R$ color of the galaxies.}  It can be seen
that, in the common magnitude bin of the two samples, dS0s from Virgo
and Coma, have similar scale-lengths and central surface brightness of
the exponential component. Figure 7 also shows that fainter dS0
galaxies show smaller scale-lengths for the exponential component. This is also true when we consider only
the more secure sample of dS0 galaxies. The
coefficients $r$ and $P$ of this relation are: $-$0.50, 0.001,
respectively. For the more reliable sample of dS0 galaxies these
coefficients are $-$0.58 and 0.001, respectively. The
central surface brightness does not depend of the absolute magnitude of
the galaxy, being $<\mu_{o,B}>=22.50$ mag. arcsec$^{-2}$, and $\sigma=0.89$. There are five galaxies (three
 from Coma and two from Virgo) which
have $\mu_{o,B} \ge 23.5$ mag arcsec$^{-2}$. They also show very large
scale-lengths for the disks. These galaxies are compatible with being 
LSB galaxies (de Blok et al. 1995). If we do not consider these five galaxies then the mean
central surface brightness of the galaxies is
$<\mu_{o,B}> =22.24$ mag. arcsec$^{-2}$, and $\sigma=0.58$. If we consider the more reliable sample of
dS0 galaxies in Coma and the Virgo galaxies without the LSBs, then we
obtain:  $<\mu_{o,B}> =22.43$ mag. arcsec$^{-2}$, and
$\sigma=0.79$. The fact that the central surface brightness of the
exponential profiles of dS0 galaxies does not depend on the luminosity
of the galaxy is similar to what happened with the disks of bright
spirals (Freeman's law). This could indicate that the exponential
components of dS0 galaxies are disk-like structures similar to those
 shown by bright spirals.

Binggeli \& Popescu (1995) determined
the mean ellipticity of dE and dS0 galaxies in Virgo as
$0.30\pm 0.02$ and $0.37\pm0.06$, respectively. These numbers are in very good agreement with the mean
ellipticity for dEs and dS0s in Coma obtained from our data:
$0.26\pm0.02$\footnote{Unless it stated to the contrary, all the
  errors showed in this paper in the value of all mean quantities are the errors of
the mean. They were computed as the mean value of the quantity divided
by the square root of the number of elements.}, and $0.34\pm0.03$, respectively. The more reliable
sample of dS0 has a mean ellipticity of $0.36\pm0.05$. Thus, dS0
galaxies in the Coma cluster, as  was the case for those in Virgo,
are flatter objects than dEs.

\subsection{Comparison of dE/dS0 with bright galaxies in Coma}

Comparison of the structural parameters of the dEs and
bright Es in the Virgo and Fornax clusters have been used in the past to
argue about the different or same formation processes of these two
systems of galaxies. Kormendy (1985) found a discontinuity between the
structural parameters of dEs and bright E galaxies in Virgo and Leo, 
concluding that dEs are very different from the sequence of bright E
galaxies. A decade later, Jerjen \& Binggeli (1997) observed a
continuous relation between the Sersic profile shape parameter and
the absolute magnitude from dE to bright E galaxies in the Virgo
cluster. Based on this relation, they concluded that dEs are the
low luminosity version of bright E galaxies. Graham \& Guzm\'an (2003)
made a compilation of data from the literature for dE and bright E
galaxies from Virgo, Fornax, Leo and  16 dEs from the Coma
cluster. They concluded that the discontinuity observed by Kormendy (1985)
in the structural
parameters of dE and bright E galaxies was caused by the selection
effect of the sample and found a continuous  and smooth change in the structural parameters from
dE to bright E galaxies, which suggests a common physical formation processes.

Figure 8 shows the relations of the structural parameters of
dE and bright E galaxies (from Aguerri et al.\ 2004) in the Coma
cluster. Those relations are obtained with one of the largest (93
galaxies from Coma) and most homogeneous (all the structural
parameters have been obtained in the same way) samples available so far in the
literature.  It can be seen that  dE
 galaxies exhibit fainter $\mu_{o}$ and  smaller values of $r_{e}$ and $n$ than
 bright E galaxies. The two galaxies which are outside
 the relations are the two cD galaxies located in
Coma. If we remove the two cD galaxies from the relations, the Pearson coefficients of  $r_{\rm e}-M_{R}$, $n-M_{R}$ and
$\mu_{o}-M_{R}$ relations are $-$0.51, $-$0.58, and $-$0.65,
respectively, in all cases with $P<0.001$. These relations suggest that bright E and dE
galaxies form a continuous family of objects with similar physical formation processes.
 
We have also computed the mean ellipticity
of bright elliptical galaxies in Coma from Aguerri et al. (2004),
as $0.25\pm 0.02$. This shows that dE and bright E
galaxies in the Coma cluster have similar ellipticities. Figure 9 shows the cumulative distribution function of the ellipticities of
 dE, dS0, and E galaxies in the Coma cluster. The cumulative
 distribution function of dEs and Es is very similar for galaxies with
 ellipticities less than 0.3, being  different for galaxies with
 ellipticities greater than 0.3. Nevertheless, the KS test reveals that we cannot statistically exclude the possibility that 
 that Es and dEs have the same distribution function. On the contrary, the KS test
 tells us that dS0s have a different distribution function of ellipticities from E and
 dE galaxies.

We have classified as dS0 galaxies those objects with surface brightness
profiles fitted with two components (see Section 3): one Sersic
profile which dominates the innermost regions of the surface brightness profiles (similar to
the bulge in bright spiral galaxies) and another exponential
profile which dominates in the outermost regions (similar to the disks of the bright spirals). Because of this similarity
between dS0s and bright spirals we have compared the structural
parameters of the bulges and disks of bright spiral galaxies in Coma
(Aguerri et al. 2004) with the
Sersic and the exponential profiles of dS0s, respectively. Figures 10
and 11 show this comparison. Figure 10 shows that there is a continuous relation for
the scales of the exponential profiles of dS0s and the disks of bright
spirals: brighter galaxies
have disks with larger scale-lengths. The coefficients $r$ and $P$ of
this relation are $-$0.63 and 0.001, respectively. These coefficients
are $-$0.66 and 0.001 for the more reliable sample of dS0 galaxies. The central surface brightness
of the disks of bright spirals does not depend on the absolute
magnitude of the galaxy. This is the well known Freeman law, discovered by Freeman
(1970) for bright spiral galaxies. The mean central surface brightness of the
disks of bright galaxies in Coma obtained from our data is
$<\mu_{o,R}> =19.76$ mag arcsec$^{-2}$ and $\sigma=0.90$. Neither does the central surface brightness of the  exponential
profiles of dS0 galaxies  depend on the luminosity of
the galaxy,  the mean value being $<\mu_{o,R}> =20.76$ mag arcsec$^{-2}$
and $\sigma=0.99$, and
$<\mu_{o,R}>=20.39$ mag arcsec$^{-2}$, and $\sigma=0.80$ for the more reliable sample of dS0 galaxies (see
Fig. 10). Both values differ by only $1\sigma$. The structural parameters of
the Sersic profiles of dS0s and bulges of bright Coma galaxies as a
function of the $R$-band absolute magnitude of the galaxies are shows in Figure 11. It can
be seen that there is a correlation between the Sersic shape parameter
($n$), the effective radius ($r_{\rm e}$), and the bulge central surface
brightness $\mu_{o}$. Brighter galaxies have bulges with larger $n$,
$r_{\rm e}$, and brighter $\mu_{o}$. The Pearson coefficients of $r_{\rm e}-M_{R}$, $n-M_{R}$ and
$\mu_{o}-M_{R}$ relations are $-$0.47, $-$0.27, 0.32, respectively. The
$P$ coefficients  of the relations are: 0.001, 0.018, 0.005,
respectively. The Pearson coefficients of those relations for the more
reliable sample of dS0 are $-$0.53, $-$0.20, and 0.29, respectively. The
$P$ coefficients  of the relations are 0.001, 0.125, 0.011,
respectively.

These results show that the photometric parameters of the Sersic and exponential components fitted to
the surface brightness profiles of dS0 galaxies follow a continuous
relation with the photometric parameters of bulges and disks
of bright spirals in Coma, respectively. This suggests that the Sersic
and exponential photometrical components of the dS0 galaxies are
similar bulge-like and disk-like structures to those seen in bright spirals. Similarly  to what happens with dE and bright E
galaxies, we propose  dS0 galaxies as
the low luminosity version of bright galaxies.

\section{Discussion}

We have analyzed the structural parameters of a sample of dwarf
galaxies in the Coma cluster that were classified 
according to the decomposition of their surface brightness
profiles. Galaxies called dEs are those whose surface
brightness profiles are well fitted with a single Sersic profile. We
have termed as dS0 galaxies those dwarf galaxies for whose
surface brightness profiles are well fitted with two components (Sersic+exponential). All of these
galaxies have colors  $B-R \ge 1.25$, typical colors for an
evolved and old stellar population.

The selection of the galaxies by the B-R color does not ensure
    to have a sample free of contamination by spiral
    galaxies, as spiral galaxies can also have red colors (Driver et
    al. 1994). Moreover, there is a $\approx 3$ magnitude overlap of the
    type-specific luminosity functions of S0, Sp and dE/dS0 (Binggeli
    et al. 1988). A visual inspection of the images of the objects has
    confirmed the presence of one possible spiral (GMP 2914). This
    object has weak spiral arms, and a strong twist of the
    isophotes. Nevertheless, this object shows a very red color
    ($B-R=1.84$). The presence of only one object as a possible spiral
    galaxy confirms that the contamination of the sample of dS0
    galaxies is not very high, and a selection criteria including
    visual inspection and measurement of the isophotal twist is
    enought to discard spirals which might contaminate the sample.

As pointed by many authors for the Fornax and Virgo dwarf galaxies, we
have found a continuous relation of the structural parameters for the
dE and bright E galaxies in the Coma cluster.  This could 
indicate that 
dE and bright E galaxies have similar physical formation
processes. However, the common relation of the 
structural parameters of dE and bright E galaxies has been explained by
some authors as more due to the uniformity of their dark matter halos
(Navarro et al. 1997) rather than to a common formation
mechanism. Moreover, it has also been argued in the literature that the similarity of
the structural parameters of dI and dE galaxies is also indicative of
a common origin between dE and dI galaxies (Lin \& Faber 1983). This has
been recently reinforced by the finding of substantial amount of
rotation for some dE  galaxies in Virgo, similar to that shown by dIs (van
Zee et al.\ 2004a). It has been proposed that dEs would be evolved dI galaxies that have lost their gas due
to ram pressure stripping with the hot intracluster medium (van Zee et
al. 2004a).  Although there is much discussion in the literature about
the origin of dEs, little  is mentioned about dS0s. Usually, dE and dS0
galaxies are taken as galaxies belonging to the same family of
objects, those with an old and evolved stellar population. But are they really the same class of objects? 
What is the origin of dS0 galaxies? Have dS0s and dEs similar
formation processes? In order to answer these questions we should
compare the structural and kinematic properties of dS0 and dE
galaxies in the Coma cluster. 

\subsection{Comparison between dE and dS0 galaxies in the Coma cluster}

\subsubsection{Photometric parameters}

First, we have compared the typical scales of dS0s and dEs in the Coma
cluster. We  computed the effective radius of all dwarf galaxies in
the same way. This was done from the fitted models of their surface
brightness profiles by resolving the equation $L(r_{\rm e})=L_{T}/2$,
 $L_{T}$ being the total luminosity of the model. Sersic or
exponential profiles have an infinite radial extension, while the
observed surface brightness profiles only reach  the sky
background surface brightness. Thus, we computed the total luminosity
of the fitted models with the expression $L_{T}=2 \pi \int_{0}^{r_{\rm c}} r
I(r) dr$, where $I(r)$ is the intensity of the profile, and $r_{\rm c}$
is the truncation radius. The truncation radius adopted for each
profile was the radial distance corresponding to an $R$-band surface brightness
of 24.5 mag arcsec$^{-2}$. This is the typical maximum radial extension of
our surface brightness profiles (see Figs 2 and 3). We obtained  the following
 mean $r_{\rm e}$ of dS0 and dE galaxies in
Coma: $2.32$ kpc ($\sigma=0.80$ kpc) and $2.83$ kpc ($\sigma=0.95$ kpc), respectively. The
more secure sample of dS0 show $<r_{\rm e}> =2.23$ and $\sigma=0.85$. Thus,
within the errors, dEs and dS0s in Coma have similar scales.

We have  measured the $B-r$  colors of the dwarf Coma galaxies. The
 mean  $B-r$ color  of dE  galaxies  was 1.78 $\pm$ 0.02
 (1.75 $\pm$ 0.04 for  dS0s). The two  samples of galaxies have the  same mean color
 within the errors. The KS test shows that we cannot exclude the possibility that the
 two  color  distribution of  dEs  and dS0s  comes  from  the same  color
 distribution.  We also  analyzed  the $B-r$  color  of the  dE and  dS0
 galaxies as  a function of the  position in the cluster. The dE and
 dS0 galaxies were divided  in two families: 
 those located at  $R/r_{\rm s} \leq 2$ and those with  $R/r_{\rm s} > 2$. The
 $B-r$  color of  galaxies at  $R/r_{s}  \leq 2$  was 1.79 $\pm$ 0.02  and
 1.81 $\pm$ 0.01 for  dEs and dS0s, respectively, while the  galaxies located at
 $R/r_{\rm s} > 2$ have colors of: 1.80 $\pm$ 0.03 and 1.64 $\pm$ 0.09 for dEs
 and dS0s, respectively. Figure 12 shows the color histograms of the
 galaxies. Although the mean color of dS0 galaxies located in the
 outer parts of the cluster is bluer than that for those located in the
 innermost region, the KS test shows that we cannot exclude the possibility that the color
 distribution  of those two families of dS0 galaxies comes from
 the same distribution. The same happens for dE galaxies located in
 the inner and outermost regions. 

We ran a two dimensional KS test in order to investigate the
differences in the spatial distributions between dEs and dS0s in the
cluster. The test reveals that we cannot exclude the possibility that dEs and dS0s have
the same spatial distribution in the cluster. We have also run the two
dimensional KS test for the spatial distribution of dwarfs (dEs and
dS0s) and bright galaxies (E/S0 and Sp). The results were that the
spatial distribution of dEs is statistically different from bright
galaxies (E/S0 and Sp). But, this is not the case for dS0s. We cannot
exclude the possibility that the spatial distribution of dS0 galaxies is different
from E/S0 or Sp galaxies. No differences have been found concerning the color and
 spatial distribution when we consider only the more reliable dS0
 sub-sample of galaxies.

The previous results prove that there is not a clear distinction
between dE and dS0 in Coma by scale, B-R color or spatial
distribution. This was also pointed out by Ryden et al. \ (1999) for a
sample of early-type dwarf galaxies in Virgo. They did not found any combination
of parameters from the surface photometry that statistically
correlates with the dE/dS0 designation. They only found that dS0
galaxies seem different from the dEs in the ellipticity. The dS0
family may
represent the tail of the distribution toward flatter shapes, which is
in agreement with what we pointed out in section 4.1.

\subsubsection{Kinematic properties}

The KS test showed that we cannot exclude the possibility that dEs and dS0s have the
same radial velocity distribution. One important kinematic
parameter is the velocity dispersion of each type of galaxy in a
cluster, which gives information about the orbits of the galaxies
(Biviano \& Katgert 2004). Figure 13 shows the velocity histograms for
bright (E and Sp) and dwarf (dE and dS0) galaxies. We have
studied the gaussianity of these velocity distributions by computing
their kurtosis ($K$) and skewness ($S$). A Gaussian distribution has
$K=3$ and $S=0$. We have obtained $K-3=0.62, 0.69, 0.65$, and 1.27, and
$S=0.14, 0.24, 0.08$, and 0.25 for Es, Sps, dEs, and dS0s, respectively. For
the more reliable sample of dS0 galaxies we  obtained $K-3=1.43$ and
$S=0.04$. It
can be seen that dS0 and Sp  galaxies have a velocity distribution
that deviates more from Gaussian. The others present similar $K$ and $S$. We have computed the mean velocity and the 
velocity dispersion of each type of galaxy using three different
methods: fitting a Gaussian to
the velocity histograms (see Fig. 13), directly measuring  the mean and
the 
dispersion of the velocities of each group of galaxies, and using the 
robust estimator defined by Beers et al. (1990). The results are given
in Table 3. 

The $\chi^{2}$ values of the Gaussian fits given in Table 3 tell us that Sp and dS0 galaxies show the smallest Gaussian distribution
of  velocities, as was also confirmed by the kurtosis and
skewness of their velocity distributions. This means that the results
obtained by these Gaussian fits are not very reliable. We have also directly computed
 the mean velocity and velocity dispersion of the different
groups of galaxies, and we have run statistical tests\footnote{The statistical
  tests used were the Student's $T$-statistic and the $F$-variance
  test. The first test gives the probability than two distributions
  have significantly different mean, and the second  gives the
  probability that the distributions have different variances.} in
order to find which
group of galaxies has statistically different mean velocity and/or
velocity dispersion from the others. We may conclude that we cannot
discard statistically the possibility that all groups have the same mean velocity and
velocity dispersion. Nevertheless, it should be noted that the E--dE
and Sp--dS0 groups of galaxies give probabilities greater than 0.92 in
the $F$-variance test. This means that there is a hint that the 
E--dE and Sp--dS0 groups of galaxies have similar velocity dispersions. Beers et al.\
(1990) proposed a more robust estimator for the computation of the
mean velocity and velocity dispersion that does not assume a Gaussian
distribution and is optimized for a sample with a small number of objects. It
can be seen in Table 3 that the differences in the mean velocity of
the different groups of galaxies are smaller than $2\sigma$. 
Although dS0 and Sp galaxies have slightly higher velocity dispersion than Es
and dEs, this difference is also less than $2\sigma$.  According
with these results, we can then say that there is a hint that dS0 and Sp galaxies
have slightly higher velocity dispersion than Es and dEs, but that the
difference is not statistically significant.

Biviano \& Katgert (2004) found
that late type galaxies follow orbits with anisotropic velocity
distributions.  Although it is not
statistically significant, the slightly higher velocity dispersion
observed for the group of Sp--dS0 galaxies with respect to the E--dE group
could be related to differences in the anisotropy of the
orbits. This means that Sp--dS0 galaxies would have more anisotropic orbits than those in the
E--dE group. This could be related with infalling motions of the
galaxies in the cluster (Solanes et al.\ 2001). Thus, dS0 galaxies would be objects falling
into the cluster following anisotropic orbits, while dEs would be
galaxies in isotropic orbits similar to E galaxies.


\subsection{Possible formation mechanism of dS0 galaxies}

In the previous subsection we have seen the main similarities and
differences between dS0 and dE galaxies. We can now try to answer the
question about the formation scenario of dS0 galaxies.

One possible explanation could be that dS0s would be an intermediate stage
between dI and dE galaxies. In this framework, dI galaxies lose
their gas content and cease star formation due to the gas stripping
mechanism.  The result would be a
dS0-like galaxy. Subsequent harassment interactions of the dS0
galaxy with   the  cluster potential and other galaxies could transform
the dS0 into a dE galaxy. The three types of galaxies taking part in
the present evolution (dI, dS0 and dE) have similar scales. The
physical processes driving the evolution should then imply no change in the
scales of their stellar distribution. The gas stripping would drive
the evolution  from dI to dS0 without change in the stellar
distribution because this process affects  the gas content of the
galaxy but not  its stellar content. Nevertheless, the transformation
from dS0 to dE should be made by harassment interactions, which affects
to the stellar component of the galaxies. It is known that repeated
tidal shocks suffered by a dwarf satellite galaxy can remove the kinematic
signature of a dI galaxy and transform it into a dE  (Mayer et
al. 2001a,b), but the final stellar distribution has a typical scale twice as
 short (Mayer et al.\ 2001a,b). This means
that this framework can hardly explain the similar scales observed in
the three types of objects, so dS0 galaxies are not an intermediate
step in the evolution from dI to dE galaxies, at least for the
observed objects. It could be that dS0 galaxies were stripped
dIrrs. But, the presence of two photometric components for dS0s
makes this unlikely. Usually, the surface brightness of dIrr galaxies
is well fitted with only one Sersic ($n=1$) profile (van Zee 2000).
 
In the harassment  scenario, dE and dS0 galaxies can  be formed by tidal
 interactions  of  bright disk  spiral  galaxies with  the  gravitational
 cluster  potential and  fast  galaxy--galaxy encounters  (Moore et  al.\
 1996; Mastropietro et al. 2004). Aguerri et al. (2004) found that bright spiral galaxies in the 
 Coma cluster are certaintly suffering the effects of  harassment: the
 scale-lengths of the disks of  bright spiral galaxies are shorter
 than those from similar field spirals. Nevertheless, the structural parameters
 of the bulges of bright spirals in the
 Coma cluster are not affected. If this picture is correct  and most  dE or dS0 galaxies are
 created by this  mechanism, then we can think a scene in which dS0
 galaxies would be harassed bright spirals that have not lost the
 disk component completely. In this picture, dE galaxies would be one step
 further in the harassment scene. They would be bright spirals that
 have fully lost their disks. If this model is valid, and
 assuming that the central part of the objects do not change by
 harassment, then the structural parameters  of dEs and
 the bulges of dS0s  should be similar to the  structural parameters of the
 bulges of bright spiral  galaxies.  Figure 14 shows
 the structural parameters of the dE galaxies and bulges of dS0,
 late-type, early-type and S0 galaxies in Coma. Table 4 shows
 the mean structural parameters of these types of galaxies. It can be
 seen that the scale of the inner parts of dS0 galaxies is similar to
 the bulges of late-type spirals, while dE galaxies show a larger
 scale. This result is similar if we consider only the more reliable
 sample of dS0 galaxies. This result indicates that dS0s are compatible
 with having been harassed late-type galaxies. But this is not so clear for
 dEs. In can be seen in Fig.\ 14 that the effective radii of the dE
 galaxies are in general larger than those of  bulges of
 bright spirals. Nevertheless, some of them have similar scales to those of
 the bulges of early-type or S0 bright galaxies. We have found that 60$\%$
 of dE galaxies  are located at a distance of more than
 3$\sigma$ from the linear relation $r_{\rm e}-M_{R}$ defined by the
 bulges of the bright spiral galaxies. These objects are not
 compatible with  being harassed bright spiral galaxies.

Mastropietro et al.\ (2004) have found that late-type galaxies can
evolve towards dE/dS0 systems by harassment. They conclude that
late-type galaxies do not  completely lose their disk with this
mechanism. This would be in agreement with our evolutionary picture in
which dS0 galaxies are harassed late-type spirals that have not fully
lost the disk component. Mastropietro et al. (2004) have fitted the
surface brightness profiles of their simulated dwarf 
galaxies with a single Sersic law. But in some cases the pure Sersic
profile cannot fit  the outermost points of the profile properly. These
galaxies (Gal1, Gal5 and Gal7 from Mastropietro et al.) show important
rotation features. This is
similar what we observed in the surface brightness profiles of the
dwarf galaxies in Coma. If, then,  our dS0 galaxies are harassed bright
spirals we can expect them to have a significant amount of rotation, similar to the simulated galaxies.

\section{Conclusions} 
 
We have obtained the structural parameters of the dwarf galaxies in
the 
Coma cluster with $-16 \le M_{B} \le -18$. The galaxies were classified into
two types according to the fit of their surface brightness
profiles. Those objects fitted with one component (Sersic profile) were
called  dE galaxies (56\%), and those fitted with two components
(Sersic and exponential profiles) were
called dS0s (29\%). The other 15\% correspond to
objects for which no reliable photometry could be obtained owing to the
presence of bright nearby objects. 

The structural parameters of dE and dS0 galaxies in Coma  are
similar to those shown by the same kinds of objects in the Virgo
cluster. There is a smooth and continuous relation between the
structural parameters of dEs and bright E galaxies in the Coma
cluster. On the other hand, the structural parameters of the
exponential profiles of dS0 galaxies are also related with those from
the disks of bright spiral galaxies in the Coma cluster.

The typical global scales of dE and dS0 galaxies in Coma are
similar, although dS0 galaxies are flatter than dEs. The
differences in color, mean velocity, and velocity dispersion between
the different groups of galaxies are not statistically significant. 
The scale of the bulges of late-type Coma galaxies is similar to the
scale of the bulges of dS0 galaxies. Nevertheless, the mean scale of dEs is
larger than that of the bulges of  all groups of bright galaxies.

Assuming  the picture in which  dwarf galaxies come from harassed bright disk
ones, we  conclude that dS0 galaxies are harassed late-type spirals
that have lost much of their disks. On the contrary, most of
the dE
galaxies (60$\%$) do not fit well in this evolutionary model, and
other physical mechanisms are required for their genesis.

\begin{acknowledgements}
We would like to thank to B. M. Poggianti and C. Mastropietro for useful
discussions in the preparation of this manuscript. We wish also to
thank to the anonymous referee for useful comments, which have
improved the quality of this paper. This research
has made use of the Isaac Newton Telescope operated by the Isaac Newton group
on La Palma at the Spanish del Roque de los Muchachos Observatory of
the Instituto de Astrof\'\i sica de Canarias. The authors were founded by
the Spanish DGES, grant AYA2001-3939.

\end{acknowledgements}

 \clearpage 
 
\begin{figure*} 
\plotone{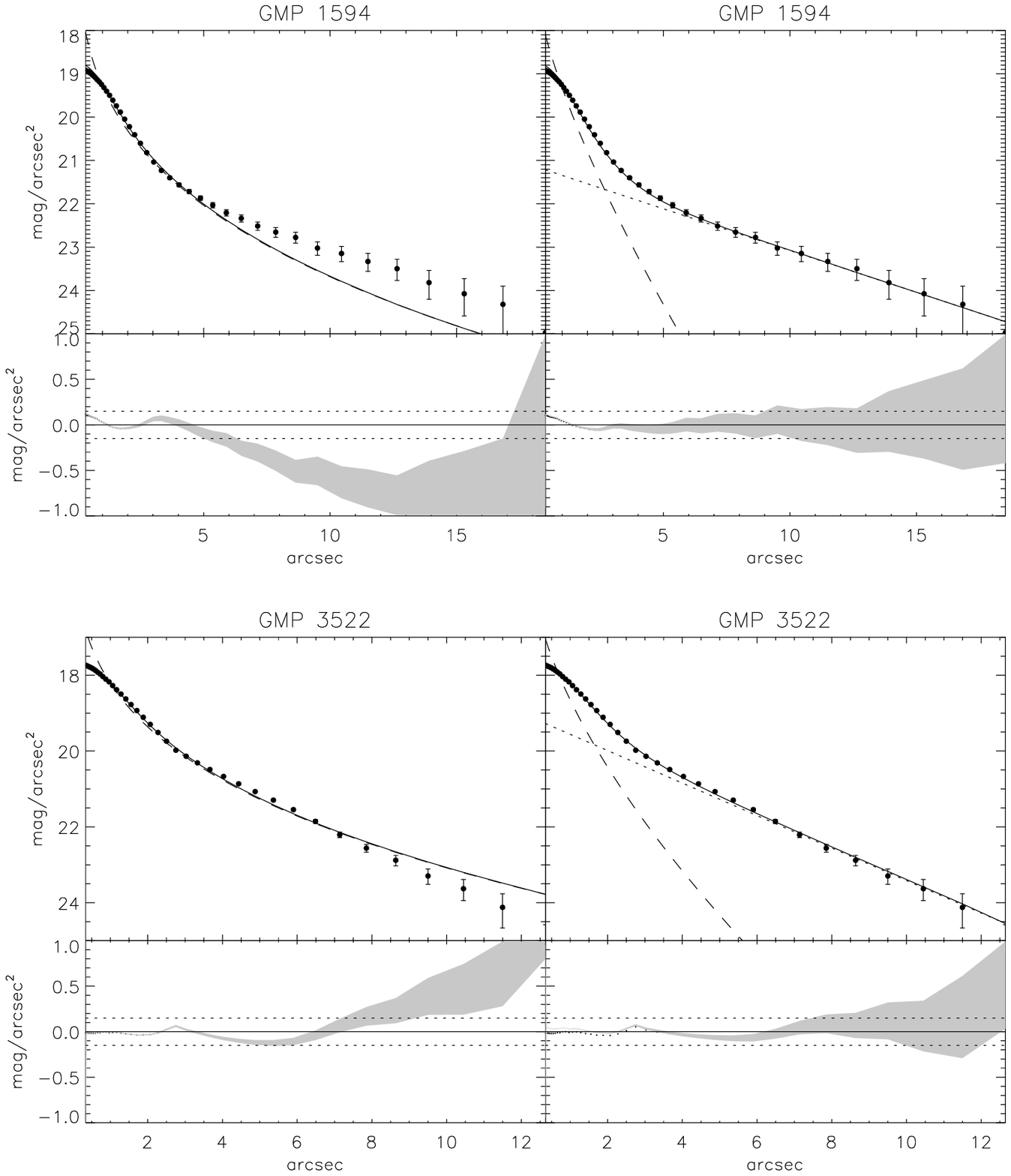}
\caption{Surface brightness profiles of GMP 1594 and GMP 3522 fitted
  with one Sersic profile (left panels) and Sersic+exponential
  profiles (right panels). The dashed line represent the Sersic profile, the dotted line the
  exponential one. The full line represents the convolved total fitted
  model. Also shown are the residuals of each fit. The dotted
  horizontal lines represent residuals of $\pm 0.15$ mag arcsec$^{-2}$. } 
\end{figure*}

\clearpage 
 
\begin{figure*} 
\plotone{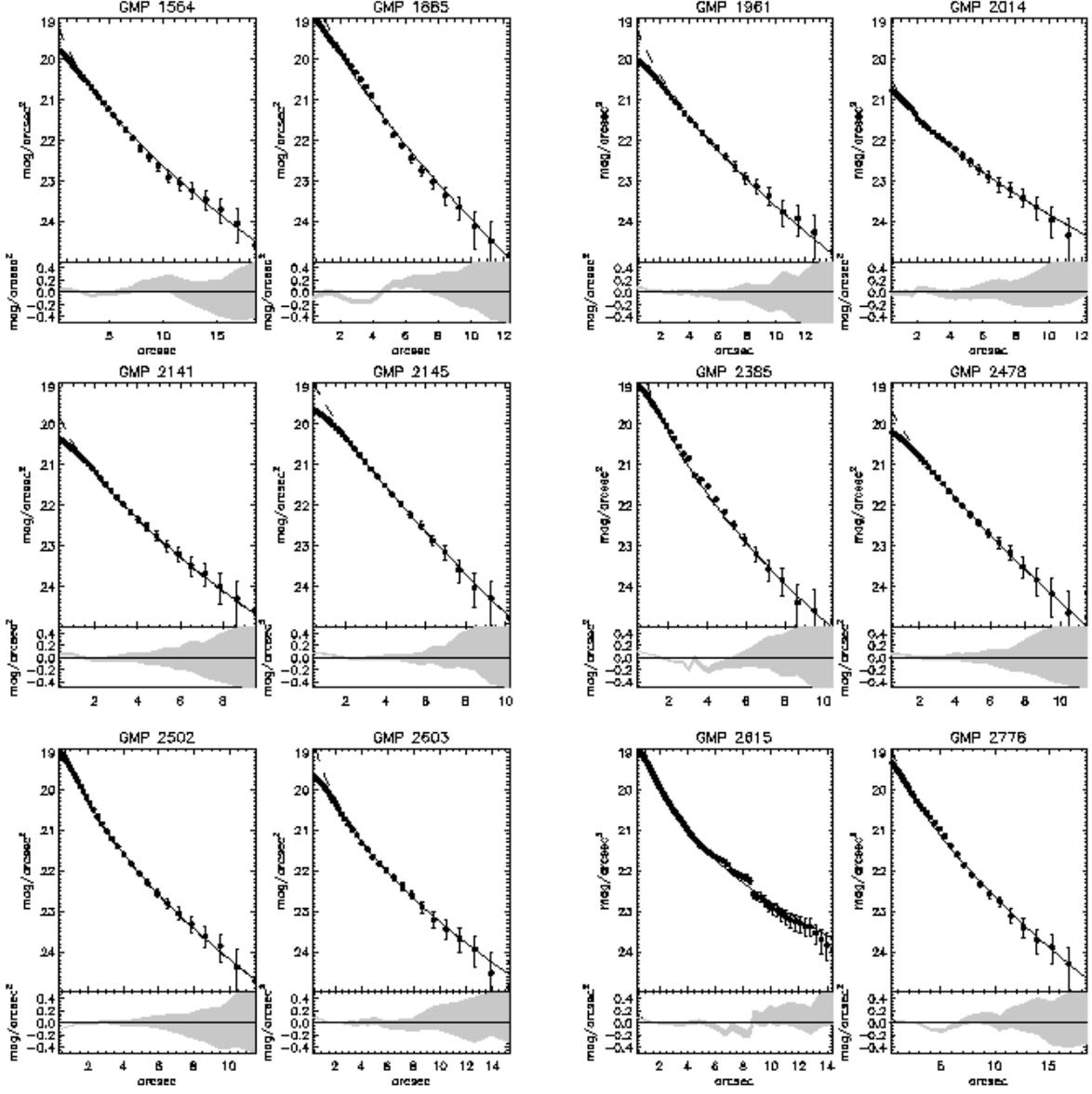} 
\caption{Surface brightness, ellipticity, and position angle isophotal profiles
 of the dE galaxies. Also overplotted is the Sersic profile fitted (dashed
 line) to the surface brightness profiles and the residuals. The full line
 represents the  convolved Sersic fitted profile. } 
\end{figure*} 
 
\clearpage 
 
\addtocounter{figure}{-1} 
 
\begin{figure*} 
\plotone{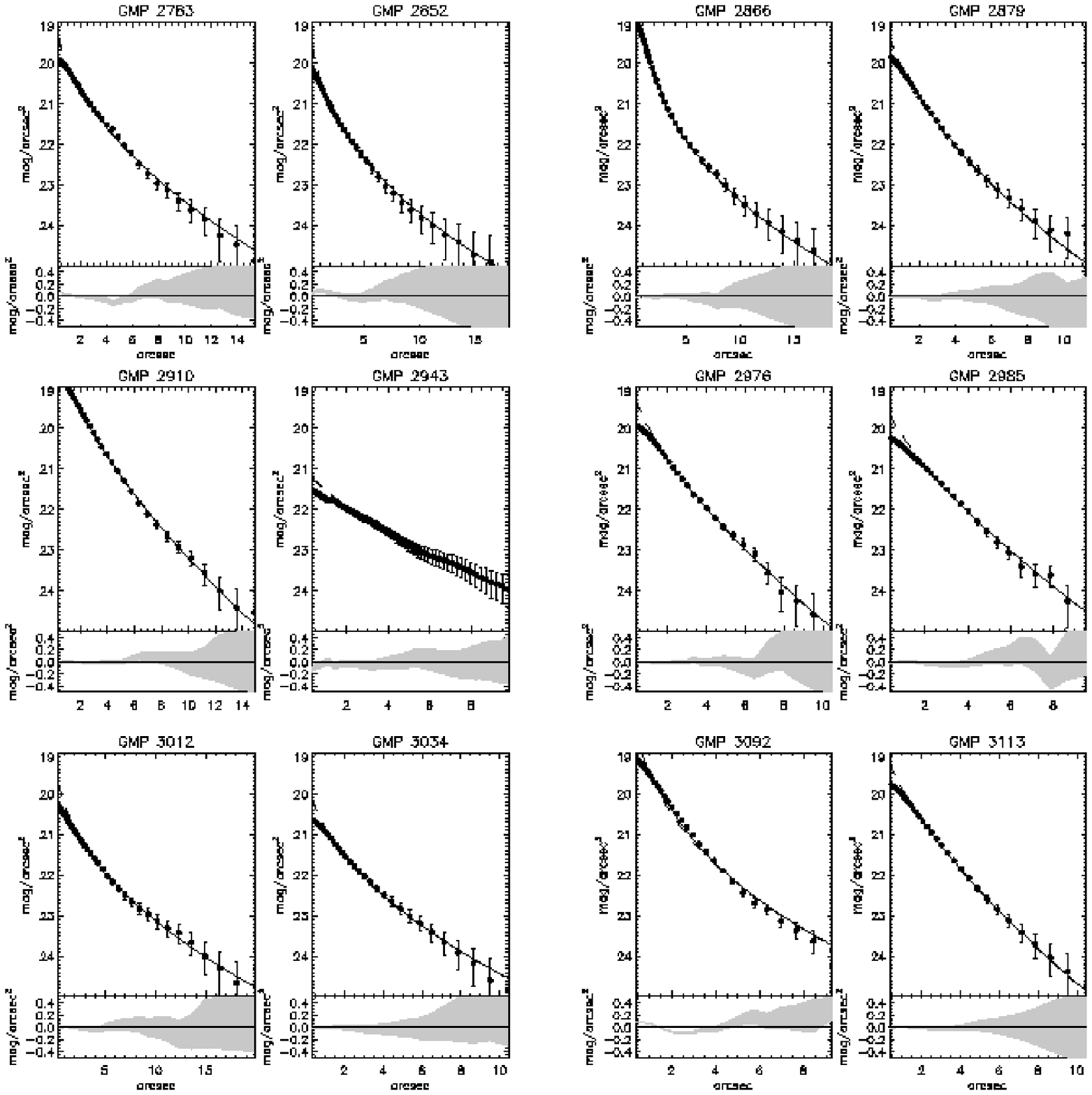} 
\caption{Continued} 
\end{figure*} 
 
\clearpage 
 
\addtocounter{figure}{-1} 
 
\begin{figure*} 
\plotone{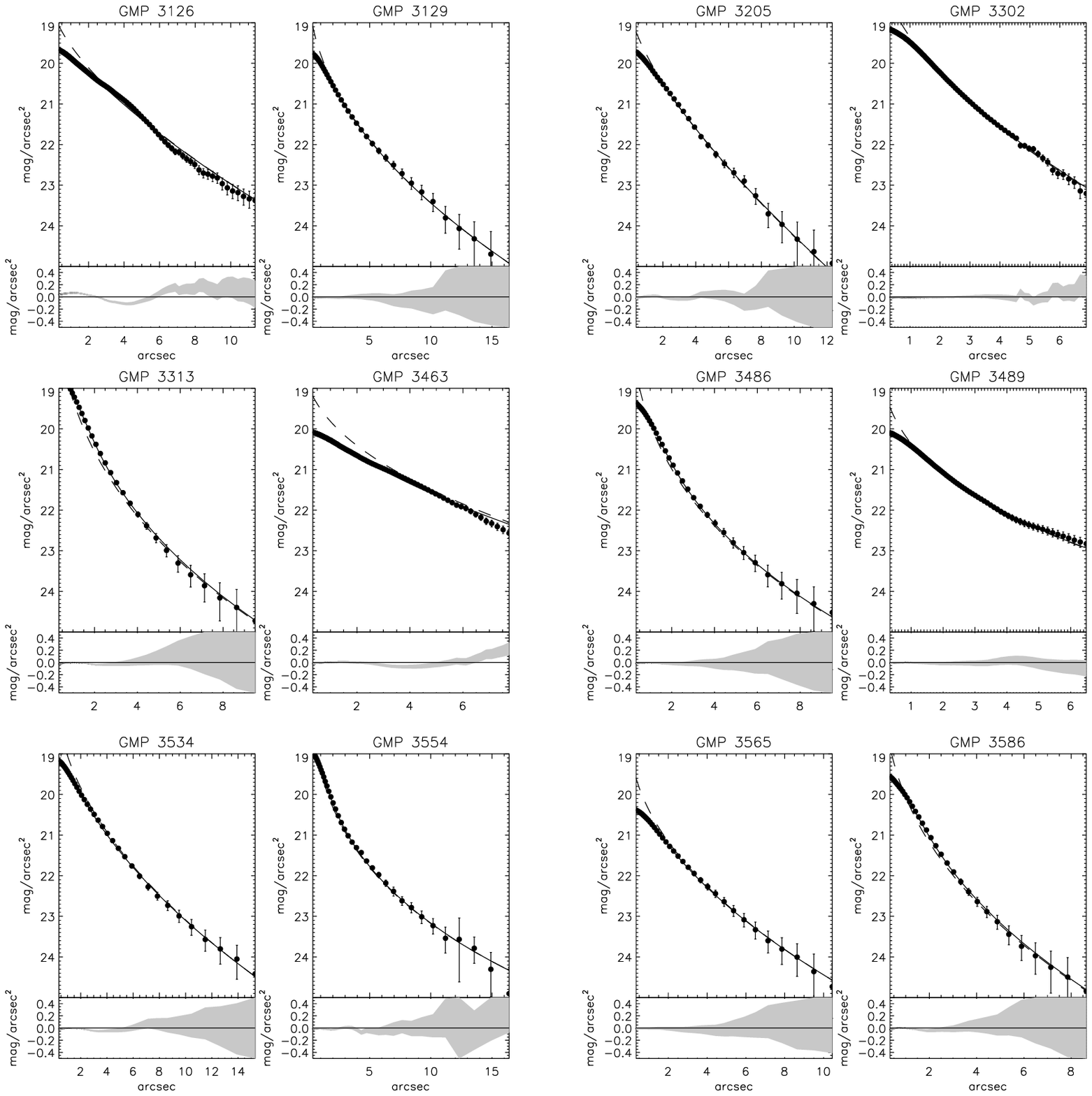} 
\caption{Continued} 
\end{figure*} 
 
\clearpage 
 
\addtocounter{figure}{-1} 
 
\begin{figure*} 
\plotone{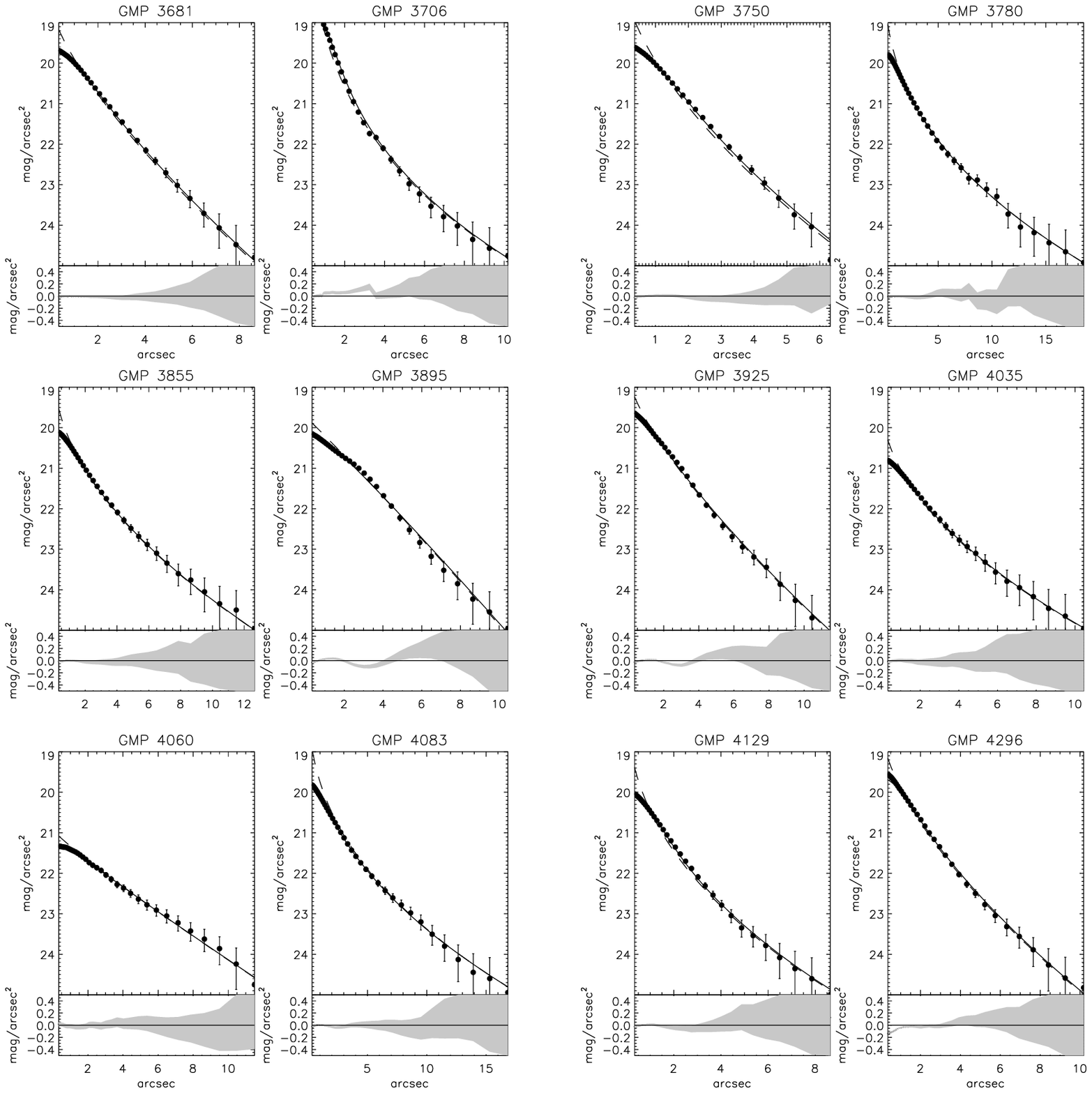} 
\caption{Continued} 
\end{figure*} 
 
\clearpage 
 
\addtocounter{figure}{-1} 
 
\begin{figure*} 
\plotone{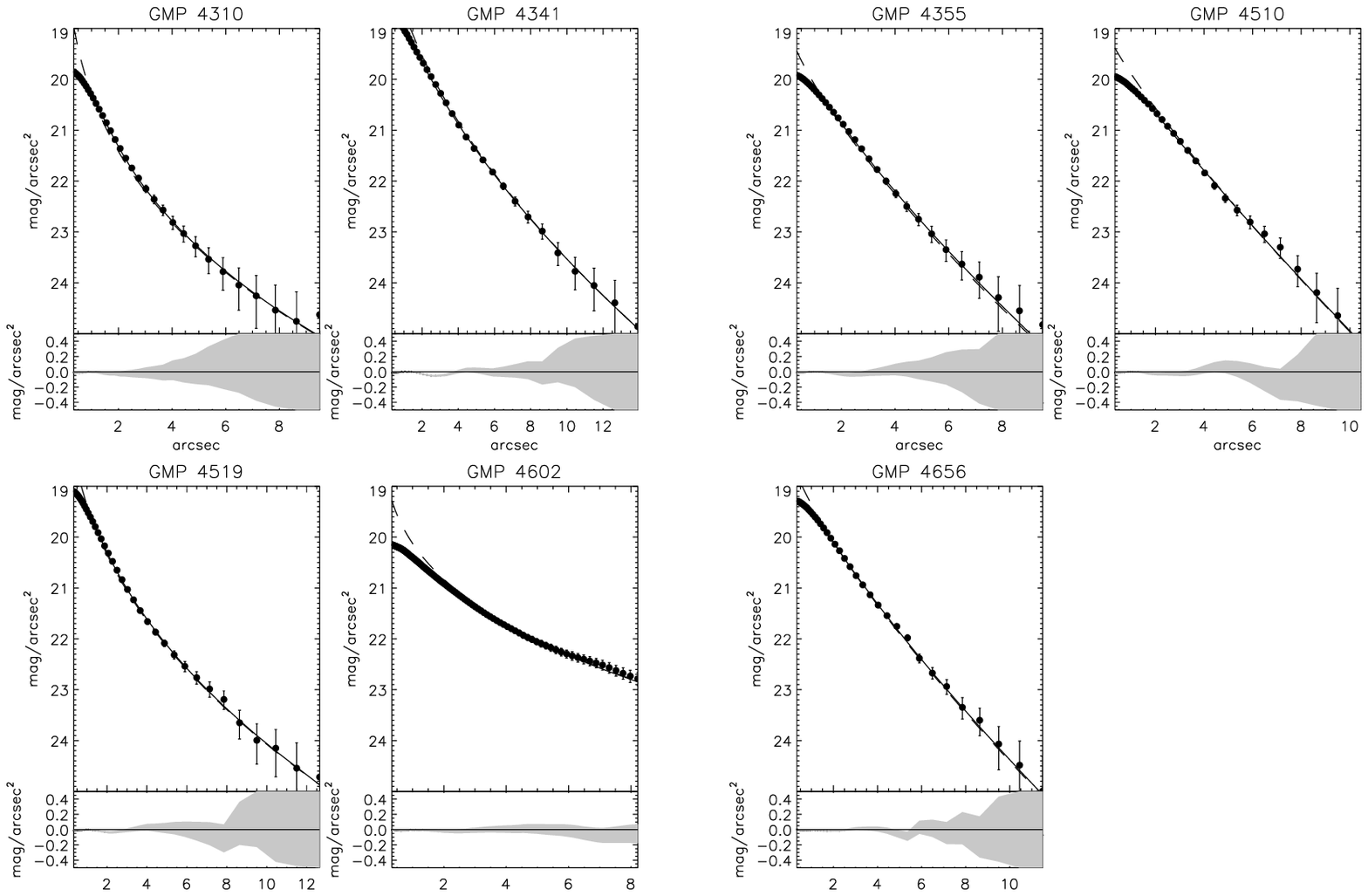} 
\caption{Continued} 
\end{figure*} 
 
\clearpage 
 
\begin{figure*} 
\plotone{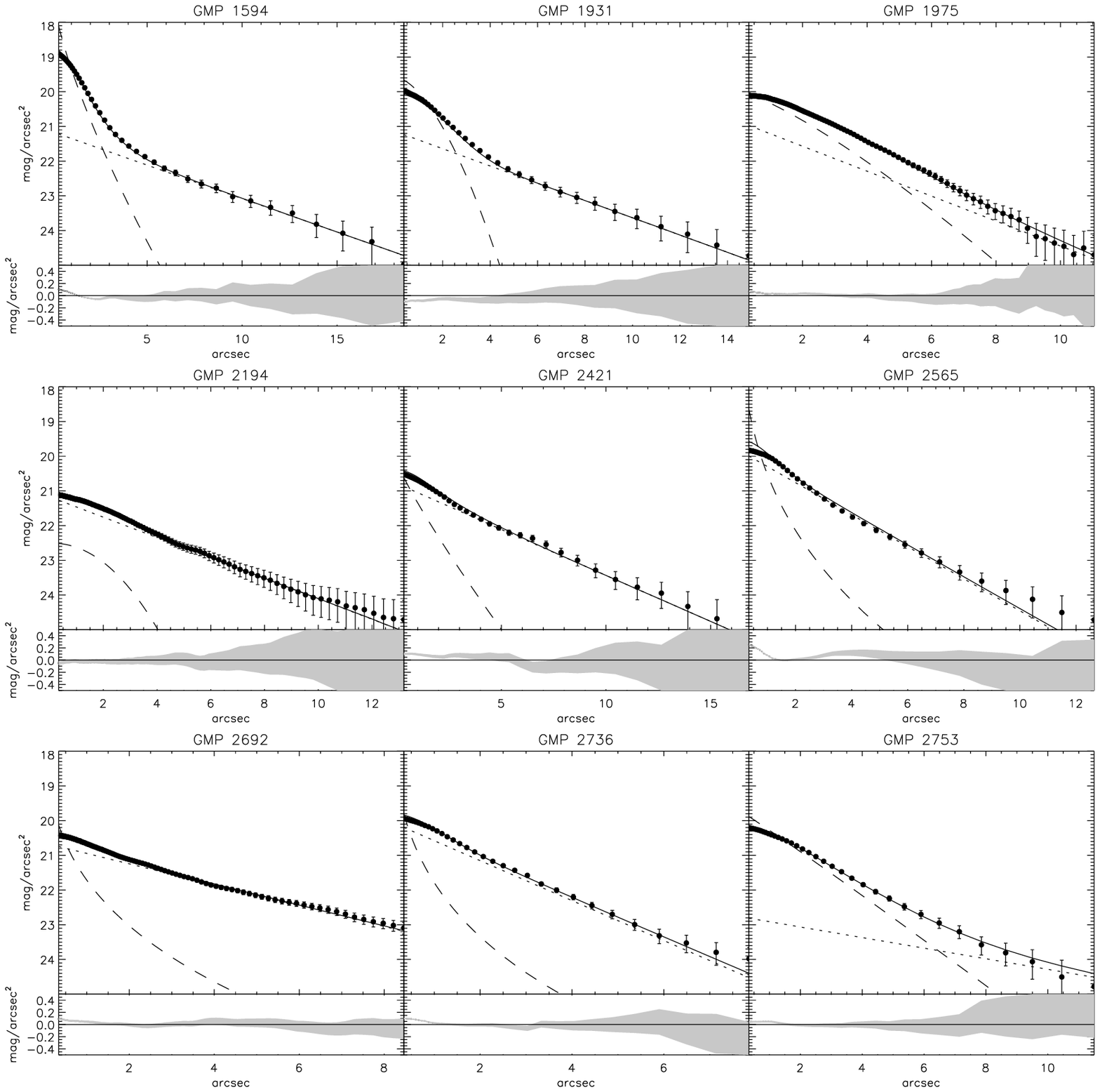} 
\caption{Surface brightness, ellipticity, and position angle isophotal profiles
 of the dS0 galaxies. Also overplotted is the Sersic (dashed line) and
 exponential (dotted line) profiles fitted to the surface brightness profiles
 and the residuals. The full line represents the total convolved fitted profile.
 } 
\end{figure*} 
 
\clearpage 
 
\addtocounter{figure}{-1} 
 
\begin{figure*} 
\plotone{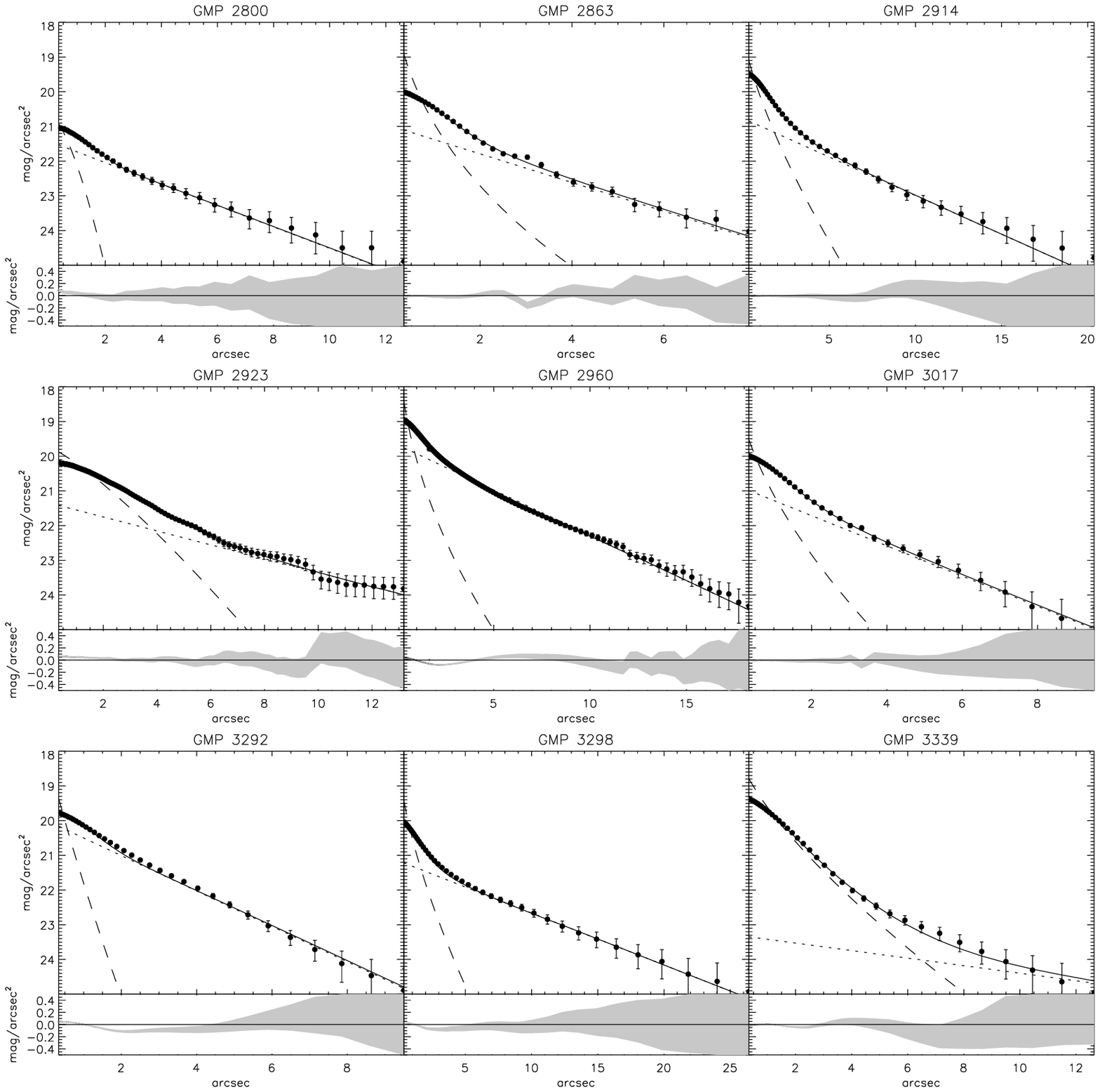} 
\caption{Continued} 
\end{figure*} 
 
\clearpage 
 
\addtocounter{figure}{-1} 
 
\begin{figure*} 
\plotone{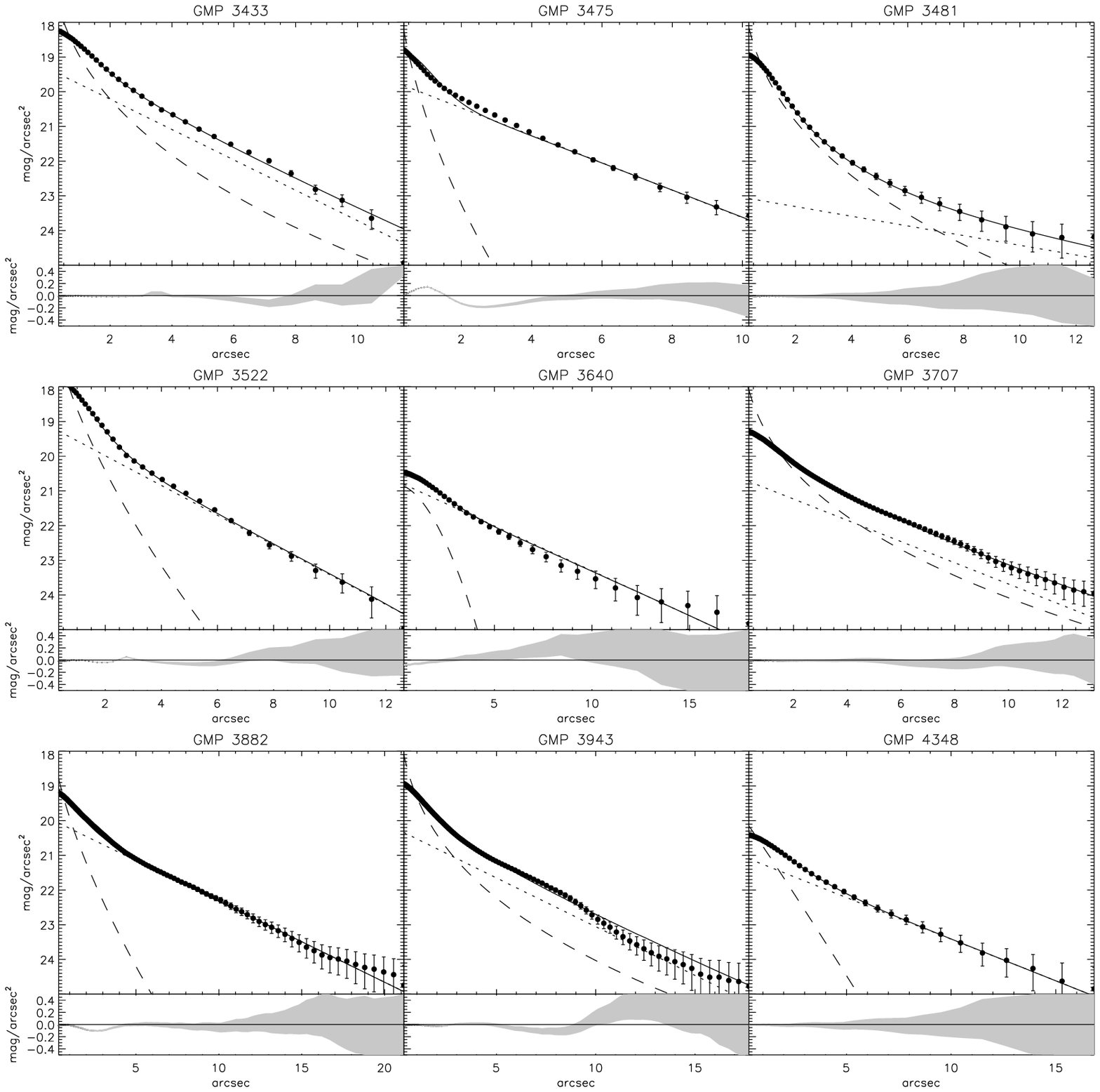} 
\caption{Continued} 
\end{figure*} 
 
\clearpage 
 
\addtocounter{figure}{-1} 
 
\begin{figure*} 
\plotone{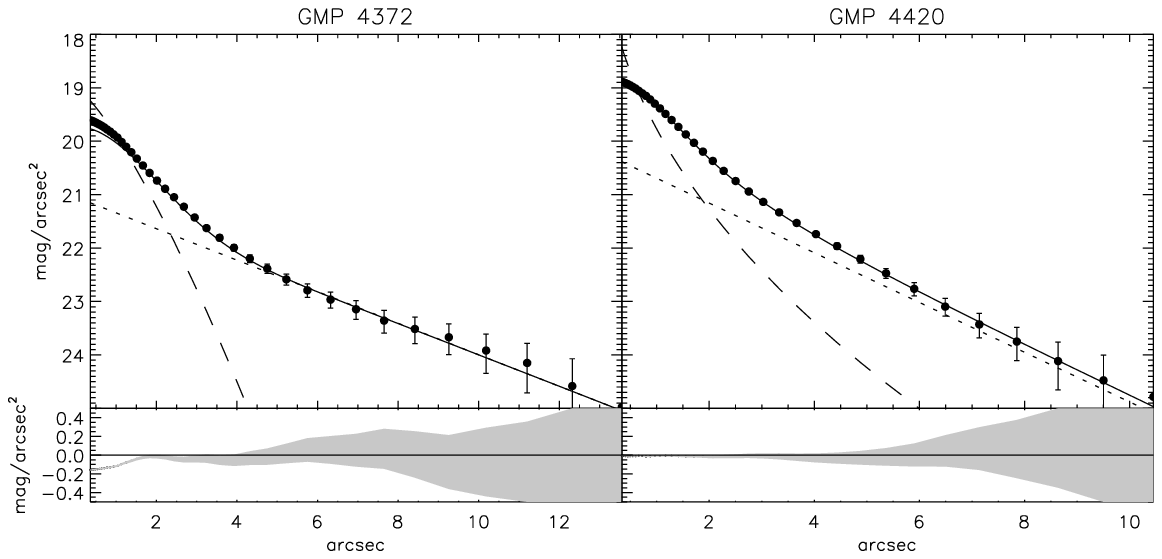} 
\caption{Continued} 
\end{figure*} 
 
\clearpage

\begin{figure*} 
\plotone{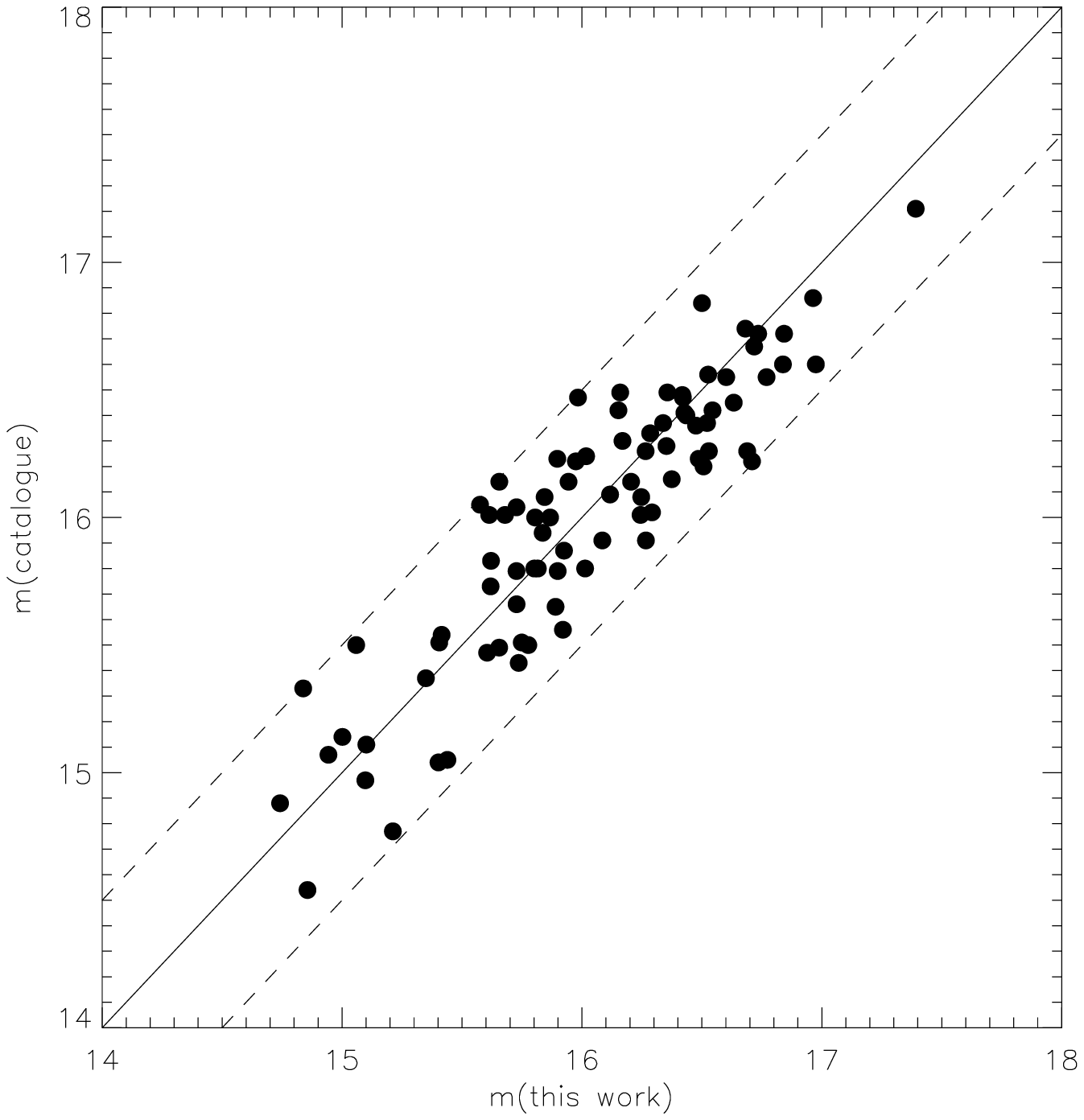} 
\caption{Comparison of the magnitudes of the objects from Godwin et al.\ (1983)
 and those 
obtained from our  model fits. The dashed lines represents a deviation of 0.5
 magnitudes.} 
\end{figure*} 

\clearpage

\begin{figure*} 
\plotone{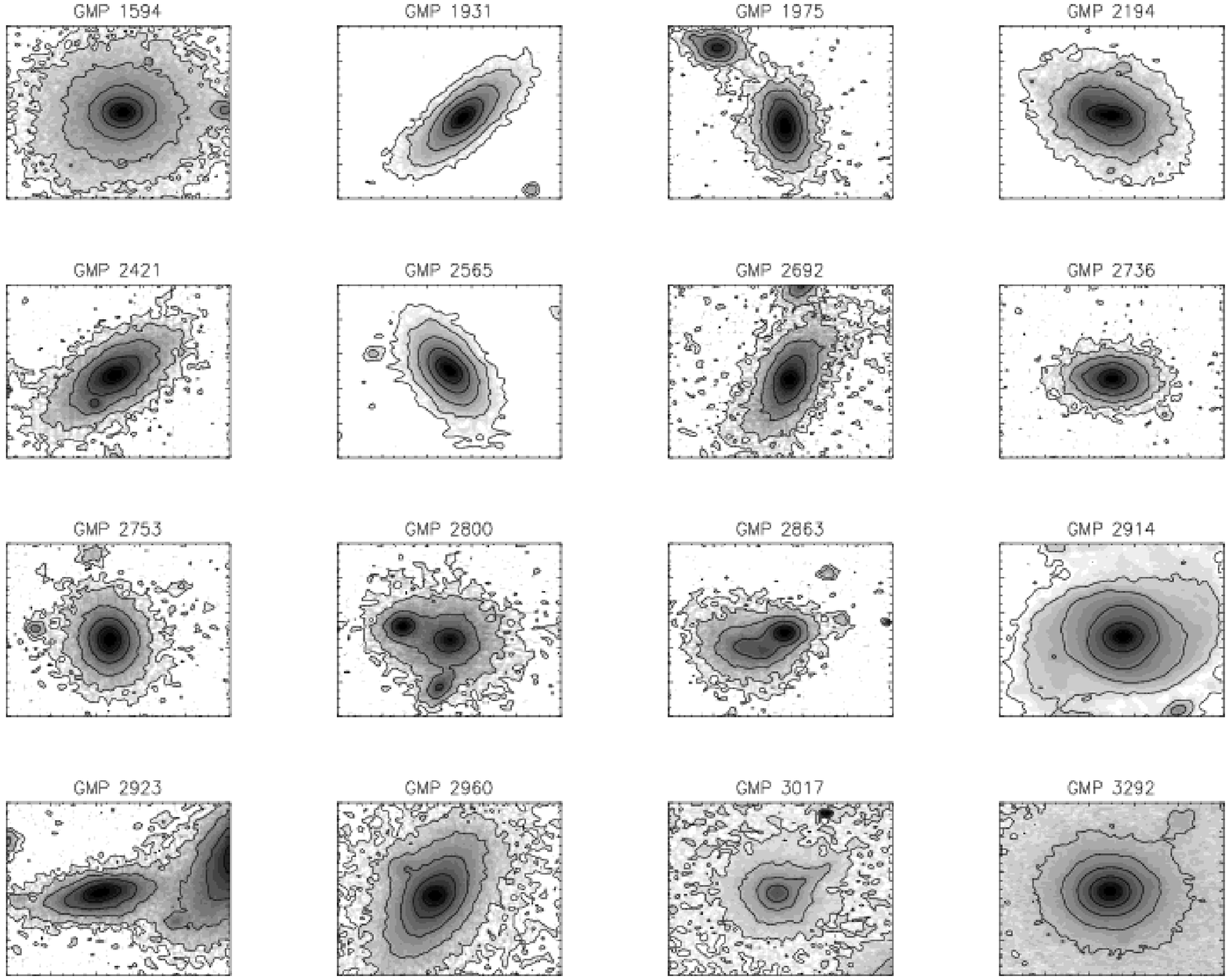} 
\caption{R-band images of the dS0 galaxies. The isocontours are spaced
  at 1 mag/arcsec$^{2}$, and the outermost corresponds to $\mu_{r}=$25
  mag/arcsec$^{2}$. The size of each panel is 33 arcsec.} 
\end{figure*} 
 
\clearpage 
 
\addtocounter{figure}{-1} 
 
\begin{figure*} 
\plotone{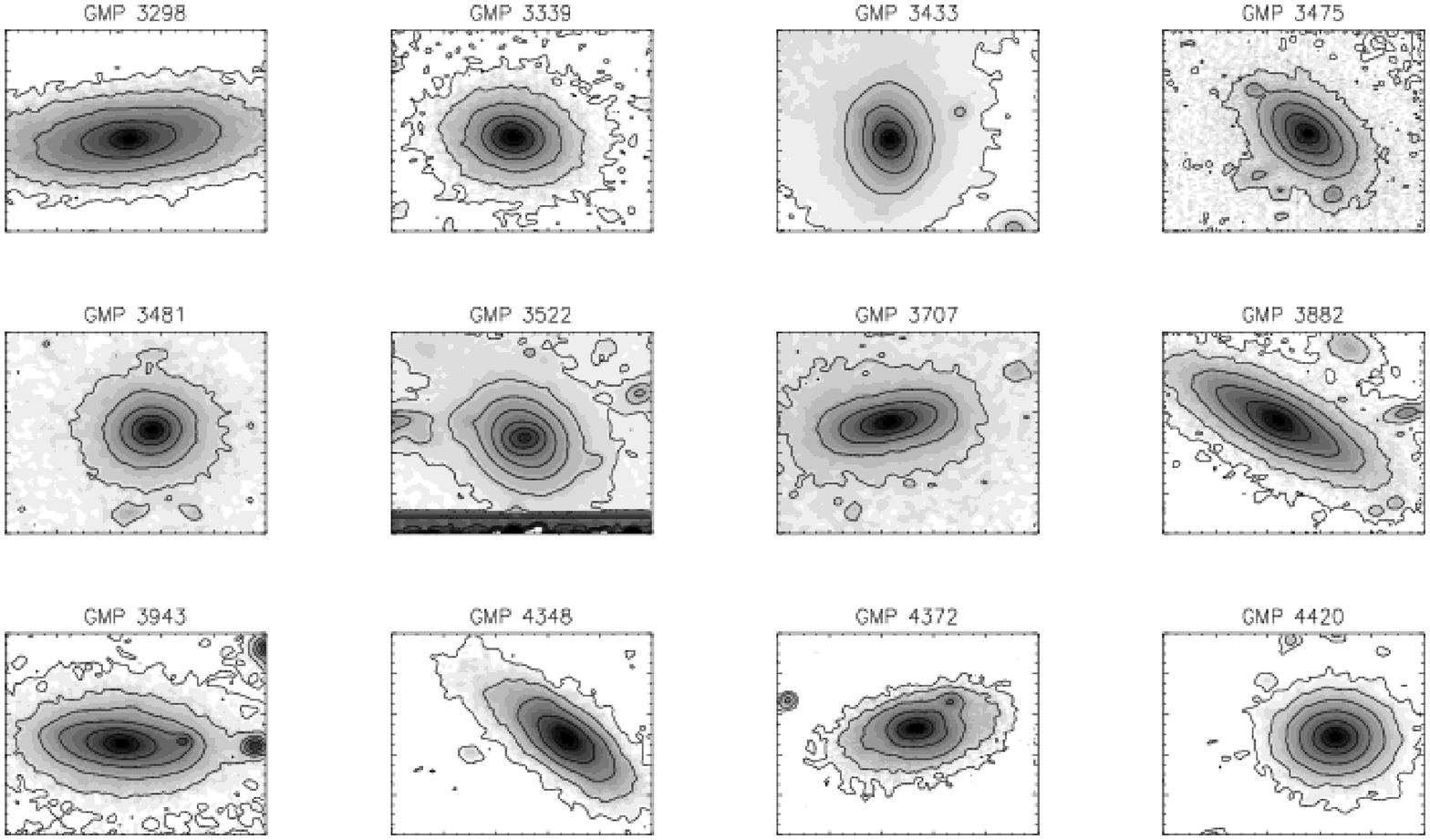} 
\caption{Continued} 
\end{figure*} 
 
\clearpage

\begin{figure*}
\plotone{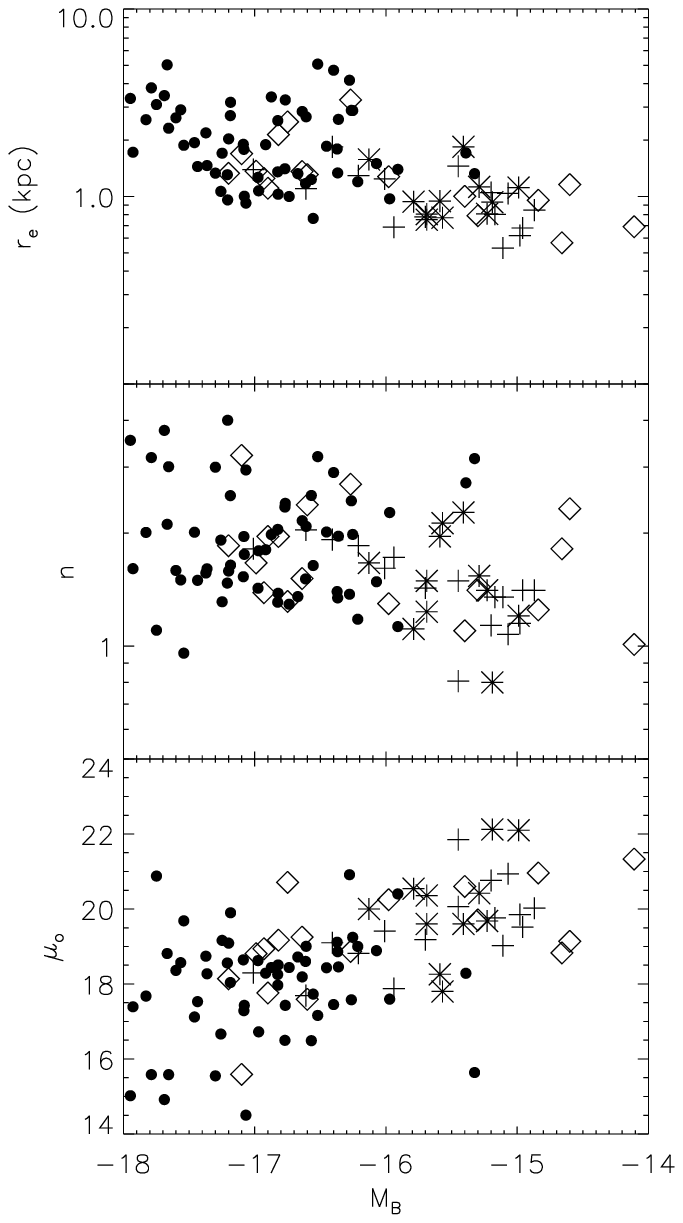}
\caption{Effective radius (top panel), shape parameter (middle panel)
  and central surface brightness (bottom panel) as a function of the
  absolute $B$ magnitudes of dE galaxies in the Coma (full points) and Virgo
  clusters. The data for the Virgo cluster are from Barazza et al. (2003)
  (diamonds); Durrell (1997) (crosses), and van Zee et al. (2004) (asterisks).}
\end{figure*}

\clearpage

\begin{figure*}
\plotone{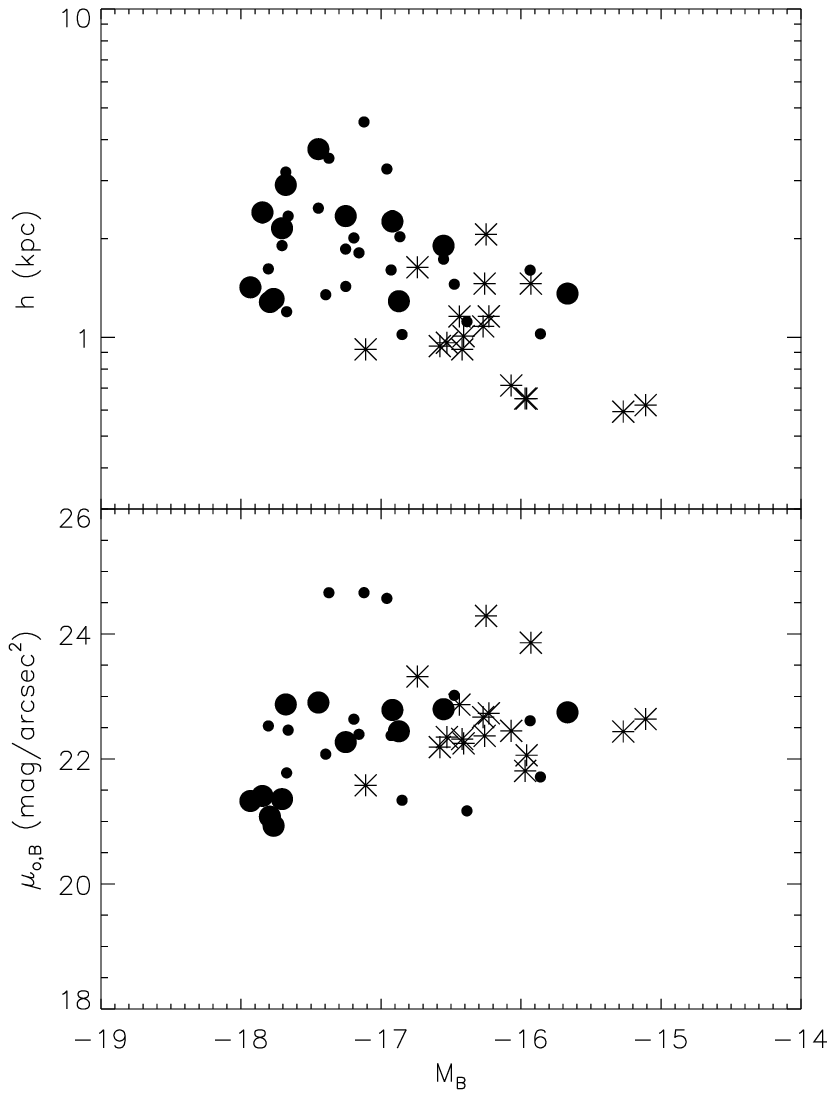}
\caption{Scale-length (top panel) and central surface brightness
  (bottom panel) of the exponential component as function of the absolute $B$-band magnitudes for the
  dS0 galaxies in the Coma (filled circles) and Virgo (asterisks). The large
  filled circles represent the dS0 Coma galaxies from the more reliable
  sample (see text for more details).}
\end{figure*}

\clearpage

\begin{figure*}
\plotone{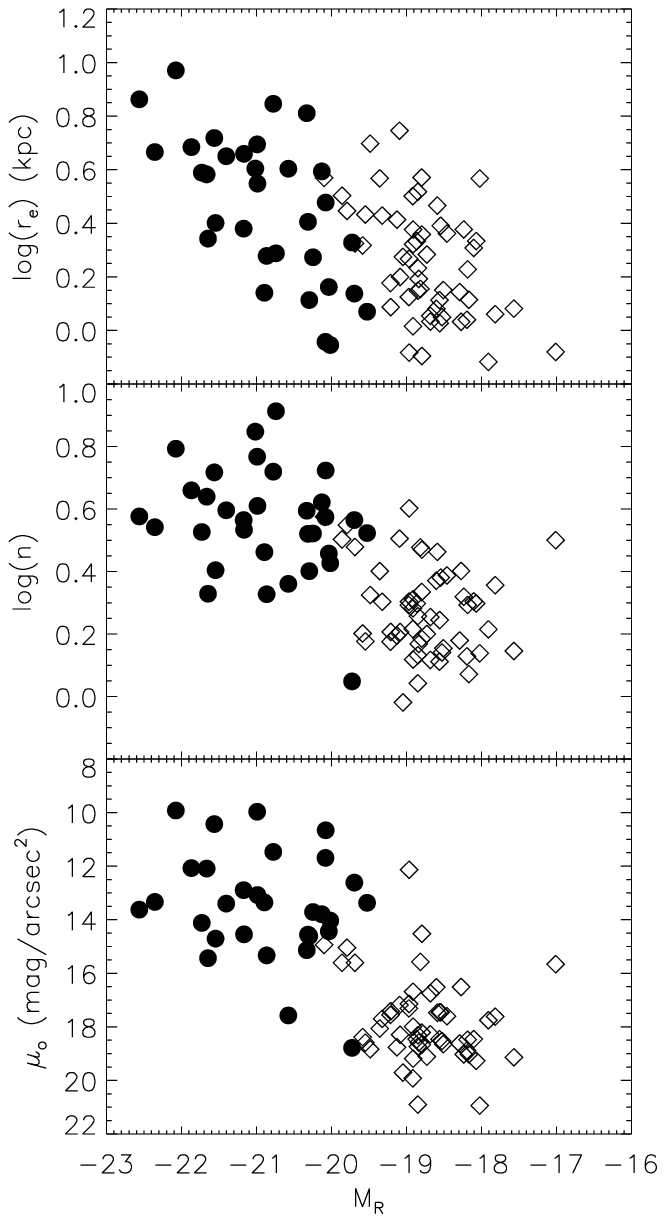}
\caption{Scale length (top panel), Sersic shape parameter (middle
  panel),  and central surface brightness
  (bottom panel) as a function of the $R$-band absolute magnitude 
  magnitudes of Bright E (full points) and dE (diamonds) galaxies in the
  Coma cluster.}
\end{figure*}

\clearpage

\begin{figure*}
\plotone{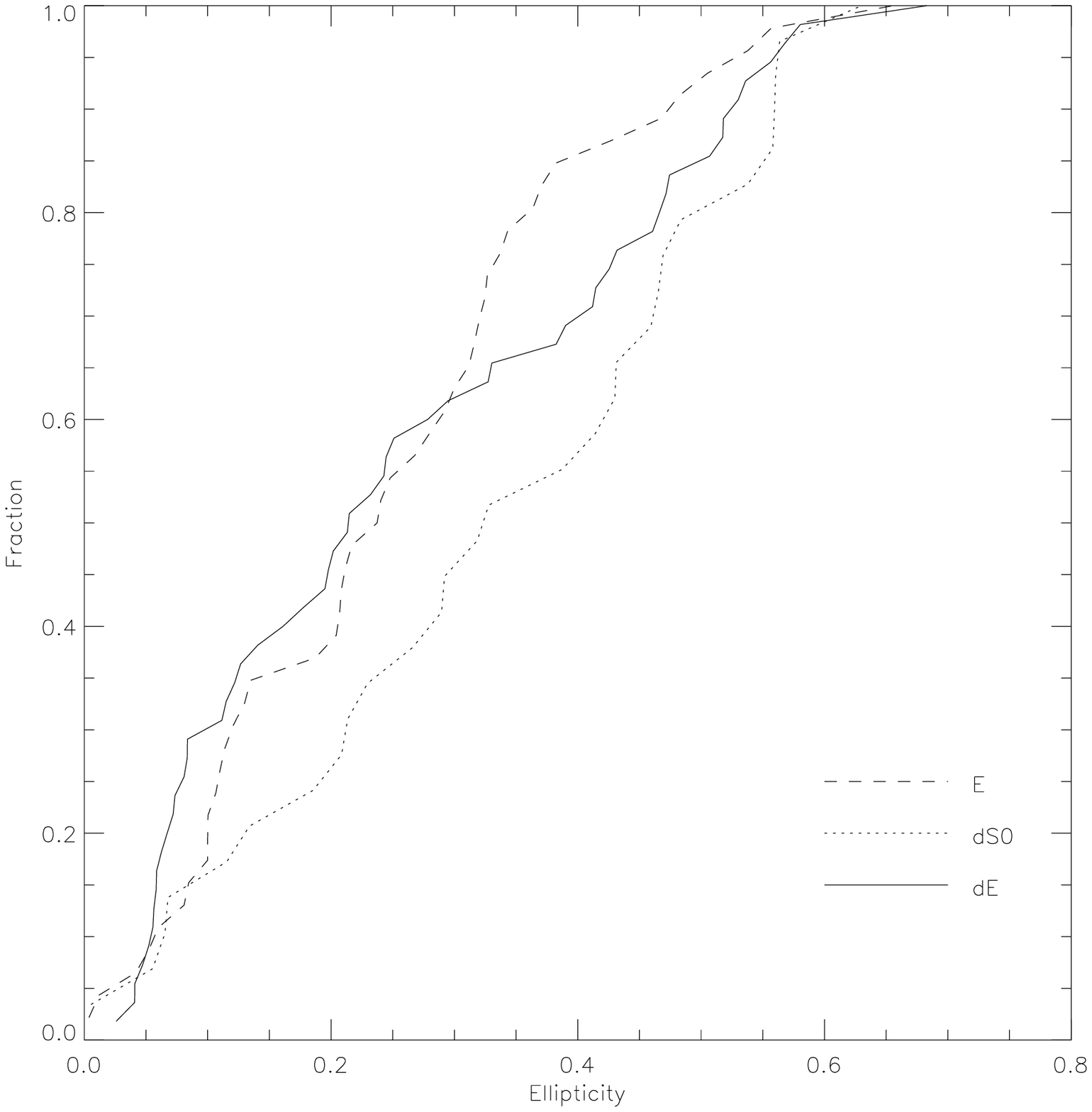}
\caption{Cumulative distribution function of the ellipticity of E (dashed line),
dE (full line), and dS0 (dotted line) galaxies.}
\end{figure*}

\clearpage

\begin{figure*}
\plotone{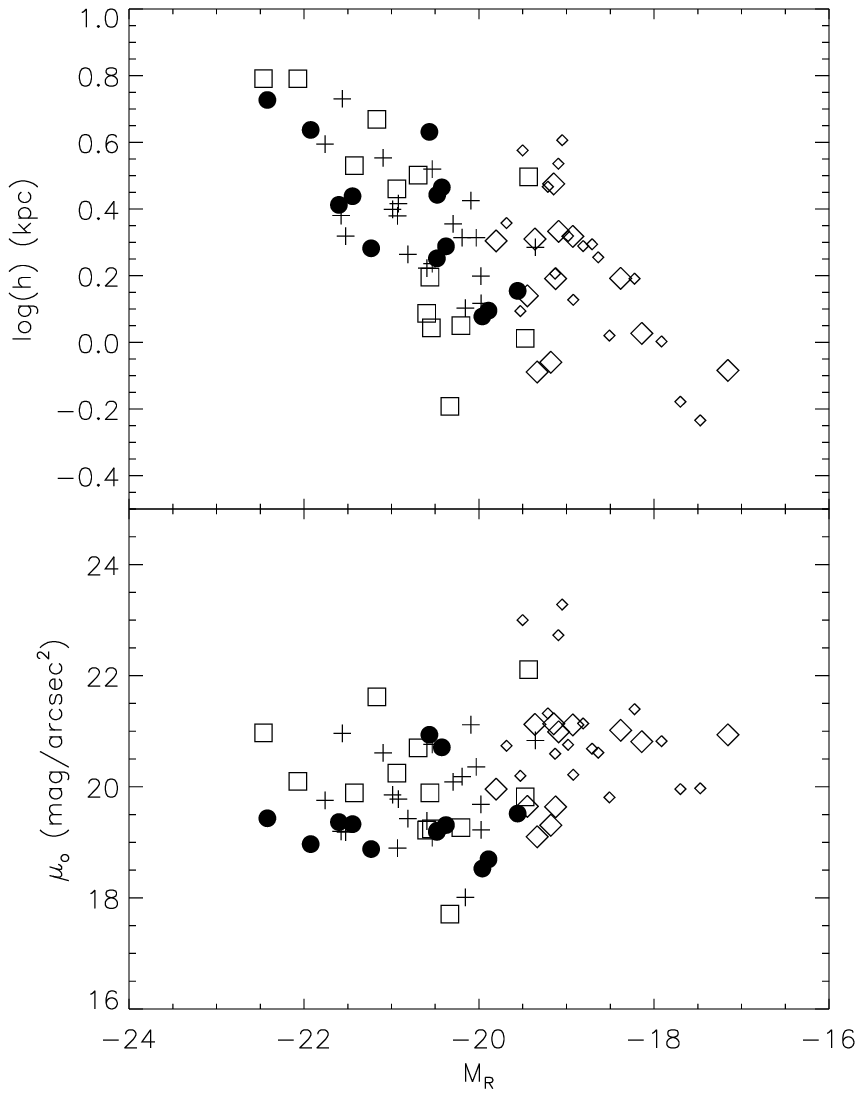}
\caption{Scale length (top panel) and central surface brightness
  (bottom panels) as a function of the absolute  $R$-band
  magnitudes of the galaxies for the disk component of dS0 and bright
  spirals in the Coma cluster. The symbols represent dS0 (diamonds),
 late type spirals (full points), early type spirals (crosses) and
  bright S0 (squares) galaxies. The large diamonds represent the more reliable
  sample of dS0 galaxies.}
\end{figure*}

\clearpage

\begin{figure*}
\plotone{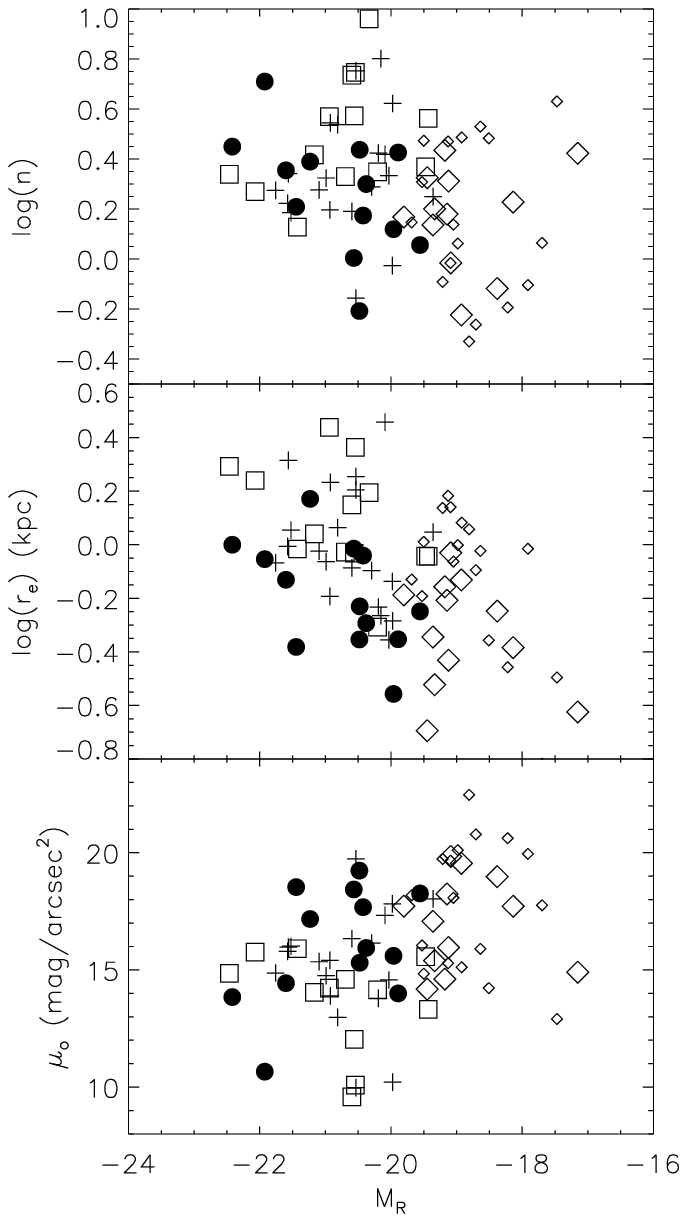}
\caption{Sersic shape parameter, $n$ (top panel), effective radius
  (middle panel), and central surface brightness (bottom panel) as a
  function of the absolute magnitude of the galaxies for the bulges of
  dS0 and bright spirals in Coma. Symbols are as in Figure 10.}
\end{figure*}

\clearpage

\begin{figure*}
\plotone{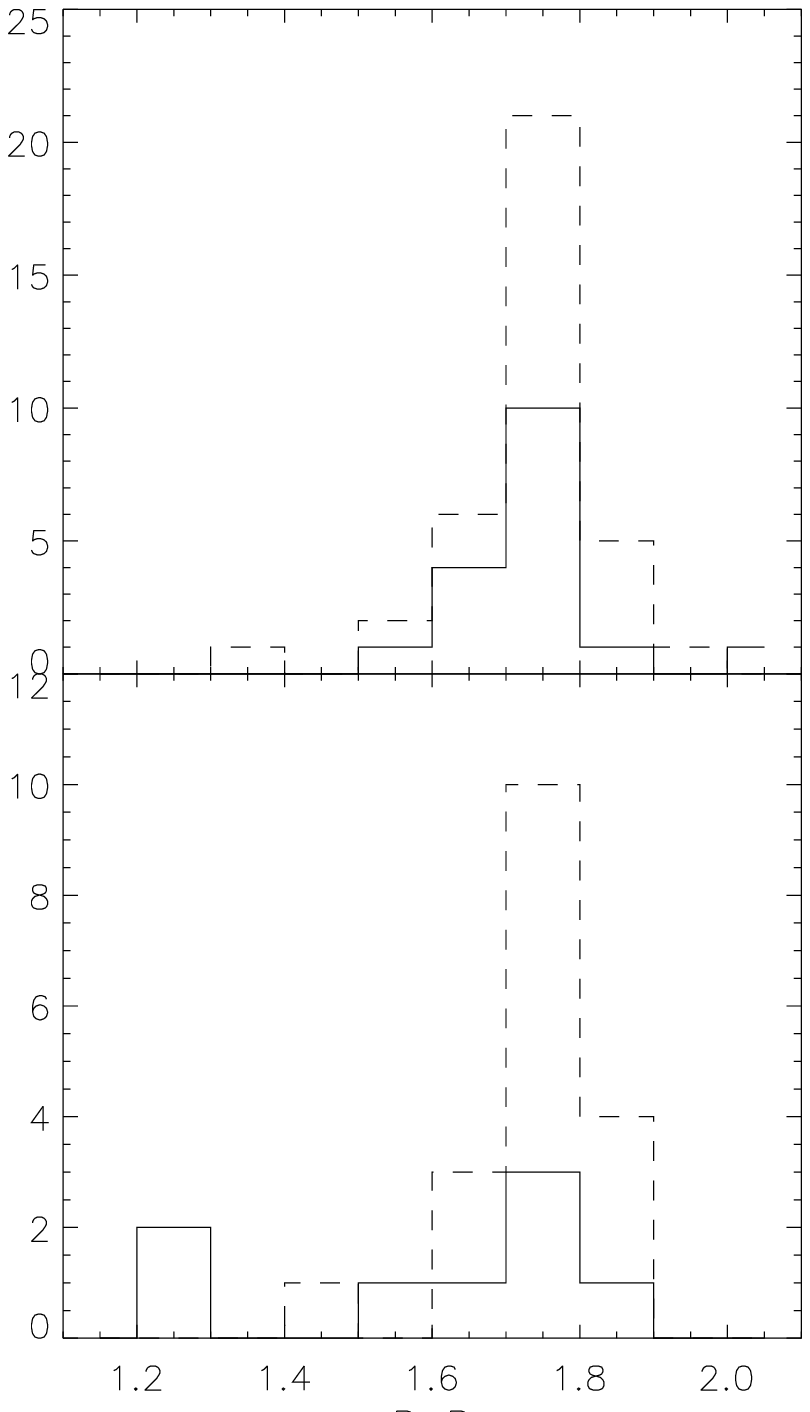}
\caption{$B-R$ color distribution of dS0 (dashed lines) and dE (solid
  lines) galaxies located at $R/r_{\rm s}<2$ (top panel) and $R/r_{\rm s}>2$
  (bottom panel).}
\end{figure*}

\clearpage

\begin{figure*}
\plotone{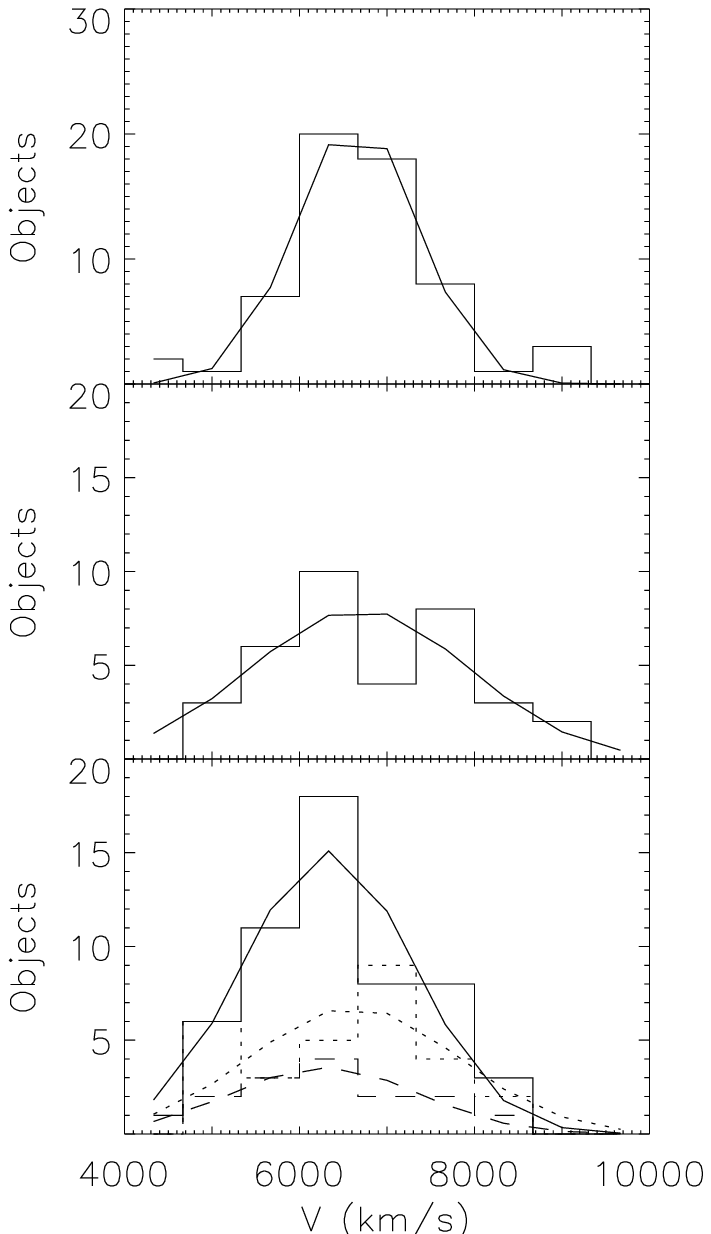}
\caption{Histograms of the line of sight velocities of ellipticals (top panel),
  bright spirals (middle panel), and dwarfs (bottom panel). In the
  bottom panel the full lines represent dEs, the dotted lines show
  dS0s, and the dashed lines correspond to the more reliable
  sample of dS0s. The Gaussian fits  of the
  different velocity distributions are also overplotted.}
\end{figure*}

\clearpage

\begin{figure*}
\plotone{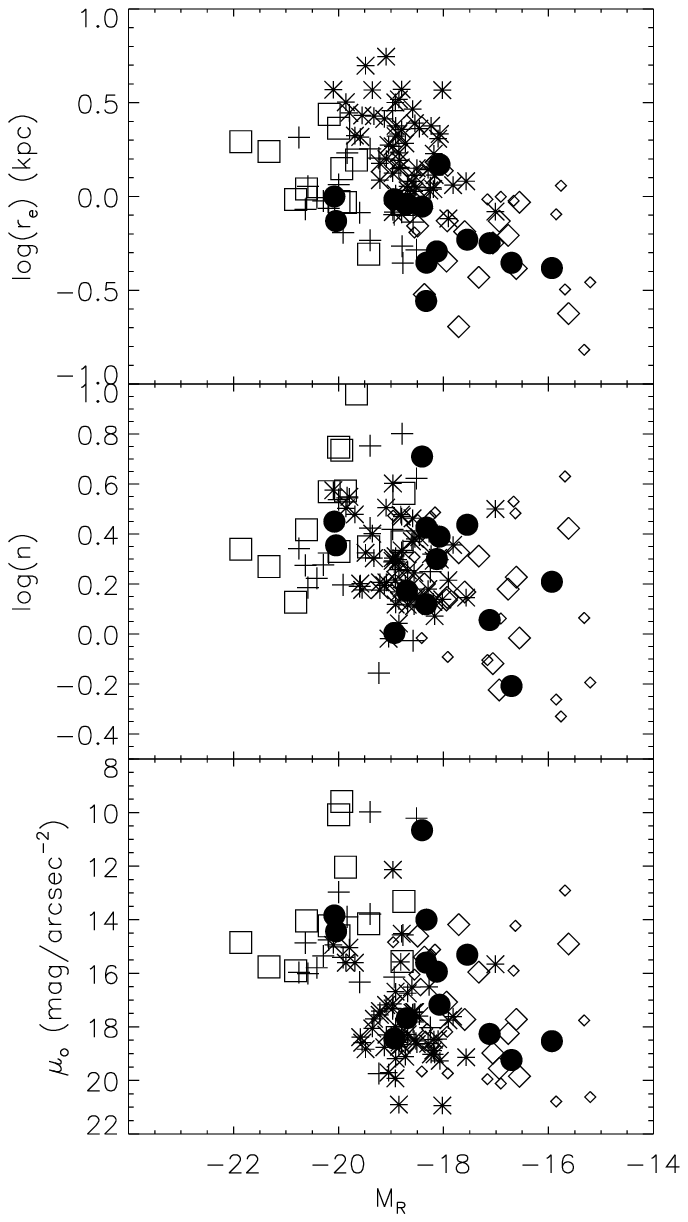}
\caption{Effective radius (top panel), Sersic shape parameter, $n$
  (middle panel), and central surface brightness (bottom panel) as a
  function of the absolute magnitude of the dE galaxies (asterisks) and  bulges of
  dS0 (diamonds), late-type (full points), early-type (crosses) and S0
  (squares) galaxies in Coma. }
\end{figure*}

\clearpage

\begin{deluxetable}{cccccccc} 
\tabletypesize{\tiny} 
\tablecaption{Structural parameters of the dE galaxies\label{tbl-1}} 
\tablewidth{0pt} 
\tablehead{ 
\colhead{GMP} & \colhead{$V_{\odot}$} & \colhead{$M_{B}$}& \colhead{$M_{R}$} &
 \colhead{$\mu_{\rm e}$} & \colhead{$r_{\rm e}$} & \colhead{$n$}& \colhead{$\epsilon$}\\
 
(1) & (2) & (3)& (4) & (5) & (6) & (7)& (8)} 
 
\startdata 
     1564&     5937&   -17.41&   -19.13&    21.81    &     6.91&     1.57&     0.47
       \\
     1885&     7802&   -17.49&   -19.21&    20.45    &     3.07&     1.50&     0.52
       \\
     1961&     7915&   -17.14&   -18.80&    21.63    &     4.59&     1.53&     0.58
       \\
     2014&     6885&   -17.23&   -18.92&    23.13    &     7.31&     1.64&     0.05
       \\
     2141&     8220&   -17.25&   -18.73&    22.20    &     3.73&     1.59&     0.08
       \\
     2145&     6834&   -16.78&   -18.56&    20.91    &     3.01&     1.29&     0.41
       \\
     2385&     7092&   -17.31&   -18.91&    20.49    &     2.32&     1.92&     0.28
       \\
     2478&     8765&   -17.31&   -18.91&    21.68    &     3.82&     1.31&     0.39
       \\
     2502&     6392&   -17.13&   -18.97&    21.21    &     3.30&     1.96&     0.11
       \\
     2603&     8181&   -17.88&   -19.32&    21.71    &     5.23&     2.01&     0.54
       \\
     2615&     6708&   -17.83&   -19.86&    22.16    &     7.53&     3.18&     0.33
       \\
     2778&     5410&   -17.64&   -19.59&    21.48    &     6.06&     1.59&     0.07
       \\
     2783&     5294&   -16.91&   -18.86&    22.41    &     6.42&     1.98&     0.06
       \\
     2852&     7451&   -17.24&   -19.36&    23.17    &     7.89&     2.52&     0.11
       \\
     2866&     6992&   -17.99&   -19.79&    22.36    &     6.35&     3.54&     0.06
       \\
     2879&     7387&   -16.97&   -18.85&    21.88    &     3.68&     1.81&     0.08
       \\
     2910&     5136&   -17.96&   -19.21&    20.55    &     3.75&     1.61&     0.20
       \\
     2943&     7289&   -16.33&   -18.02&    23.57    &     8.05&     1.37&     0.51
       \\
     2976&     6693&   -16.66&   -18.29&    21.55    &     3.30&     1.51&     0.33
       \\
     2985&     5291&   -16.41&   -17.57&    21.81    &     3.60&     1.40&     0.38
       \\
     3012&     8064&   -17.72&   -19.48&    23.06    &     9.85&     2.11&     0.47
       \\
     3034&     6439&   -16.65&   -18.24&    23.20    &     5.86&     2.09&     0.14
       \\
     3092&     8247&   -17.71&   -19.69&    21.78    &     4.08&     3.01&     0.07
       \\
     3113&     7601&   -17.26&   -18.84&    21.42    &     3.26&     1.47&     0.21
       \\
     3126&     7905&   -17.62&   -19.55&    21.49    &     5.46&     1.50&     0.47
       \\
     3129&     6729&   -16.87&   -18.91&    22.08    &     5.61&     2.05&     0.43
       \\
     3205&     6196&   -17.02&   -18.51&    21.39    &     3.62&     1.43&     0.30
       \\
     3302&     5714&   -17.12&   -18.56&    20.91    &     2.94&     1.76&     0.23
       \\
     3313&     6231&   -17.11&   -18.79&    20.57    &     2.04&     2.95&     0.04
       \\
     3463&     6618&   -16.68&   -18.80&    22.54    &     8.92&     2.16&     0.68
       \\
     3486&     7604&   -17.35&   -18.81&    21.73    &     2.99&     3.00&     0.13
       \\
     3489&     5507&   -16.44&   -18.59&    23.42    &     8.39&     2.90&     0.16
       \\
     3534&     6411&   -17.50&   -18.97&    21.15    &     4.57&     2.01&     0.57
       \\
     3554&     7125&   -17.74&   -20.10&    22.74    &     8.27&     3.76&     0.03
       \\
     3565&     7140&   -16.50&   -18.10&    22.47    &     4.52&     2.01&     0.46
       \\
     3586&     6681&   -16.61&   -18.27&    21.63    &     2.56&     2.52&     0.07
       \\
     3681&     6942&   -16.87&   -18.53&    21.18    &     2.56&     1.38&     0.06
       \\
     3706&     6892&   -17.25&   -18.96&    20.46    &     1.90&     4.00&     0.18
       \\
     3750&     6339&   -16.60&   -17.91&    20.96    &     1.91&     1.64&     0.12
       \\
     3780&     5080&   -16.30&   -18.46&    22.54    &     7.15&     2.44&     0.41
       \\

     3855&     5722&   -16.40&   -18.19&    22.38    &     4.67&     1.96&     0.20
       \\
     3895&     8535&   -17.60&   -19.05&    21.43    &     3.52&     0.96&     0.21
       \\
     3925&     6448&   -16.72&   -18.85&    21.33    &     3.46&     1.36&     0.06
       \\
     4035&     6665&   -16.30&   -18.07&    23.22    &     5.13&     1.98&     0.06
       \\
     4060&     8686&   -17.81&   -18.85&    22.94    &     6.05&     1.10&     0.20
       \\
     4083&     6202&   -16.81&   -18.54&    22.31    &     6.28&     2.40&     0.56
       \\
     4129&     5991&   -16.01&   -17.82&    22.19    &     3.04&     2.27&     0.04
       \\
     4296&     8352&   -17.42&   -19.09&    21.43    &     3.02&     1.61&     0.08
       \\
     4310&     4366&   -15.35&   -17.01&    22.17    &     3.00&     3.16&     0.25
       \\
     4341&     5434&   -17.01&   -18.68&    20.29    &     3.14&     1.80&     0.53
       \\
     4355&     6205&   -16.41&   -18.19&    21.45    &     2.80&     1.34&     0.05
       \\
     4510&     6609&   -16.26&   -18.17&    21.23    &     3.12&     1.18&     0.43
       \\
     4519&     5638&   -16.81&   -18.60&    21.26    &     3.38&     2.35&     0.24
       \\
     4602&     6444&   -16.56&   -19.09&    23.78    &    13.69&     3.20&     0.52
       \\
     4656&     5809&   -16.86&   -18.68&    20.76    &     3.09&     1.31&     0.24
       \\
 \enddata 
\tablecomments{Col. (1) GMP galaxy identification from the Coma catalog of
 Godwin et al.\ (1983). Col. (2) Velocity in km/s from Godwin et al.\ (1983). Col.
 (3) Absolute $B$ magnitude from Godwin et al.\ (1983). Col. (4) Absolute $R$ magnitude
 from the surface brightness profile fits of this work. Col. (5) effective
 surface brightness in $R$ band. Col. (6) effective radius in arcsec. Col. (7)
 Sersic $n$ parameter. Col. (8) Ellipticity.} 
\end{deluxetable}

\clearpage
 
\begin{deluxetable}{ccccccccccc} 
\tabletypesize{\tiny} 
\tablecaption{Structural parameters of the dS0 galaxies\label{tbl-2}} 
\tablewidth{0pt} 
\tablehead{ 
\colhead{GMP} & \colhead{$V_{\odot}$} & \colhead{$M_{B}$}& \colhead{$M_{R}$} &
 \colhead{$\mu_{\rm e}$} & \colhead{$r_{\rm e}$} & \colhead{$n$}&
 \colhead{$\epsilon_{\rm b}$}& \colhead{$\mu_{0}$} & \colhead{$h$} &
 \colhead{$\epsilon_{\rm d}$}\\ 
(1) & (2) & (3)& (4) & (5) & (6) & (7)& (8) & (9) & (10) & (11)} 
 
\startdata 

     1594&     5742&   -17.72&   -19.36&    19.69    &     1.25&     1.37&     0.12&    21.12&     5.62&     0.06
       \\
     1931&     7599&   -16.97&   -18.93&    20.49    &     1.55&     0.60&     0.62&    21.11&     4.35&     0.32
       \\
     1975&     5272&   -15.97&   -17.91&    21.30    &     2.89&     0.79&     0.47&    20.82&     3.01&     0.47
       \\
     2194&     8426&   -17.31&   -18.81&    23.13    &     2.16&     0.47&     0.35&    21.14&     3.68&     0.21
       \\
     2421&     8126&   -17.25&   -18.98&    22.26    &     1.95&     1.15&     0.22&    20.76&     4.09&     0.48
       \\
     2565&     7095&   -16.90&   -18.51&    20.47    &     0.98&     3.04&     0.61&    19.81&     2.35&     0.46
       \\
     2692&     7955&   -16.98&   -18.64&    22.90    &     1.90&     3.39&     0.25&    20.61&     3.60&     0.56
       \\
     2736&     4890&   -15.89&   -17.47&    21.82    &     1.03&     4.27&     0.07&    19.97&     1.88&     0.21
       \\
     2753&     7704&   -17.01&   -19.09&    21.40    &     2.85&     0.96&     0.20&    22.73&     7.10&     0.29
       \\
     2800&     7020&   -16.52&   -18.22&    21.65    &     0.79&     0.64&     0.02&    21.40&     3.51&     0.07
       \\
     2863&     4950&   -15.70&   -17.16&    20.29    &     0.76&     2.65&     0.51&    20.94&     2.62&     0.33
       \\
     2914&     7447&   -17.85&   -19.69&    20.87    &     1.58&     1.40&     0.07&    20.74&     4.87&     0.23
       \\
     2923&     8664&   -17.72&   -19.21&    21.14    &     2.53&     0.81&     0.57&    21.32&     5.40&     0.56
       \\
     2960&     5922&   -17.75&   -19.12&    20.05    &     0.99&     2.05&     0.14&    19.64&     4.15&     0.63
       \\
     3017&     6784&   -16.92&   -18.14&    21.03    &     0.97&     1.69&     0.05&    20.81&     2.49&     0.13
       \\
     3292&     4924&   -16.42&   -17.70&    19.92    &     0.49&     1.16&     0.07&    19.96&     2.13&     0.19
       \\
     3298&     6554&   -17.49&   -19.15&    21.17    &     1.50&     1.51&     0.41&    21.13&     7.23&     0.56
       \\
     3339&     6417&   -17.16&   -19.05&    20.70    &     2.14&     1.37&     0.15&    23.28&     9.98&     0.12
       \\
     3433&     5569&   -17.83&   -19.18&    20.16    &     1.98&     2.72&     0.31&    19.31&     2.47&     0.47
       \\
     3475&     8010&   -17.99&   -19.45&    18.41    &     0.40&     2.12&     0.13&    19.65&     2.74&     0.39
       \\
     3481&     7718&   -17.42&   -19.50&    20.94    &     2.11&     2.97&     0.06&    23.00&     7.77&     0.07
       \\
     3522&     5087&   -17.80&   -19.34&    18.48    &     0.93&     1.59&     0.09&    19.10&     2.53&     0.27
       \\
     3640&     7483&   -16.92&   -18.71&    21.62    &     1.71&     0.55&     0.66&    20.69&     4.19&     0.54
       \\
     3707&     7220&   -17.21&   -19.13&    21.34    &     3.35&     2.95&     0.68&    20.59&     3.54&     0.43
       \\
     3882&     6894&   -17.89&   -19.80&    20.56    &     1.50&     1.47&     0.34&    19.96&     4.64&     0.41
       \\
     3943&     5496&   -17.43&   -18.92&    21.41    &     3.48&     3.06&     0.49&    20.22&     3.86&     0.56
       \\
     4348&     7576&   -17.30&   -19.09&    21.58    &     1.96&     0.96&     0.63&    20.99&     4.52&     0.29
       \\
     4372&     6725&   -16.60&   -18.38&    20.28    &     1.33&     0.76&     0.46&    21.02&     3.67&     0.43
       \\
     4420&     8509&   -17.73&   -19.53&    20.10    &     1.21&     2.03&     0.10&    20.20&     2.33&     0.01
       \\
 \enddata 
\tablecomments{Col. (1) GMP galaxy identification from the Coma catalog of
 Godwin et al.\ (1983). Col. (2) Velocity in km/s from Godwin et al.\ (1983). Col.
 (3) Absolute $B$ magnitude fro Godwin et al.\ (1983). Col. (4) Absolute $R$
     magnitude from the surface brightness profile fits of this work. Col. (5) effective
 surface brightness in $R$ band. Col. (6) effective radius in arcsec. Col. (7)
 Sersic $n$ parameter. Col. (8) Ellipticity of the bulge. Col. (9) central
 surface brightness of the disc in $R$ band. Col. (10) scale of the disk in
 arcsec. Col. (11) ellipticity of the disk.} 
\end{deluxetable}

\clearpage

\begin{deluxetable}{cccccccc} 
\tabletypesize{\footnotesize} 
\tablecaption{Velocity dispersion of the different types of galaxies \label{tbl-3}} 
\tablewidth{0pt} 
\tablehead{ 
\colhead{Galaxy type} & \multicolumn{3}{c}{Gaussian fit} & 
\multicolumn{2}{c}{Directly} & \multicolumn{2}{c}{Biweight
  estimator}\\
   & $<v>$ & $\sigma$ &   $\chi^{2}$ & $<v>$ & $\sigma$ &  $<v>$ & $\sigma$   \\
   & (km s$^{-1}$) & (km s$^{-1}$) & & (km s$^{-1}$) & (km s$^{-1}$) & (km
s$^{-1}$)&  (km s$^{-1}$) } 
\startdata 
E &  6655 & 693 & 2.45 &   7096 & 1040  & 7057$\pm$109 & 955$\pm$145\\
Sp&  6686 &1252 & 4.42  &  7181 & 1161& 7156$\pm$238& 1281$\pm$158\\
dE&  6329 &979  & 3.23 & 6732 & 1023 & 6709$\pm$126 & 1104$\pm$96\\
dS0& 6622 &1187  & 4.38 & 6819& 1178& 6881$\pm$268 & 1288$\pm$154  \\ 
dS0$^{*}$& 6290& 1068&0.39 &  6717& 1147 & 6729$\pm$308 & 1202$\pm$171\\
 \enddata 
\tablecomments{(*) more reliable sub-sample of dS0 (see definition in
  Section 3.1)}
\end{deluxetable}

\clearpage
 
\begin{deluxetable}{ccccccc} 
\tabletypesize{\footnotesize} 
\tablecaption{Structural parameters of the dS0 galaxies\label{tbl-2}} 
\tablewidth{0pt} 
\tablehead{ 
\colhead{} & \colhead{dE} & \colhead{E}& \colhead{dS0 (bulges)} & \colhead{Spl
 (bulges)} & \colhead{Spe (bulges)} & \colhead{S0 (bulges)} } 
\startdata 
$<\mu_{o}>$ & 17.81$\pm$0.21 & 12.87$\pm$0.46 & 17.44$\pm$0.45 & 16.08$\pm$0.68
 & 14.78$\pm$0.68 & 12.80$\pm$1.04 \\ 
$<\log(r_{\rm e})>$ & 0.26$\pm$0.03 & 0.57$\pm$0.09 & -0.21$\pm$0.05 &
-0.19$\pm$0.06 & -0.01 $\pm$0.05 & 0.10$\pm$0.06  \\ 
$<\log(n)>$ & 0.28$\pm$0.02 & 0.58$\pm$0.02 & 0.17$\pm$0.05 &
0.26$\pm$0.07 & 0.34$\pm$0.05 & 0.49$\pm$0.06 \\ 
 \enddata 
\end{deluxetable}


\begin{thebibliography}{} 

\bibitem[Aguerri, Iglesias-Paramo, Vilchez, \& Mu{\~ n}oz-Tu{\~ n}{\' 
o}n(2004)]{} Aguerri, J.~A.~L., Iglesias-Paramo, J., 
Vilchez, J.~M., \& Mu{\~ n}oz-Tu{\~ n}{\' o}n, C.\ 2004, \aj, 127,
1344

\bibitem[Aguerri \& Trujillo(2002)]{} Aguerri, 
J.~A.~L.~\& Trujillo, I.\ 2002, \mnras, 333, 633 

\bibitem[Aguerri, Balcells, \& Peletier(2001)]{} 
Aguerri, J.~A.~L., Balcells, M., \& Peletier, R.~F.\ 2001, \aap, 367,
428

\bibitem[Andreon, Davoust, \& Poulain(1997)]{} Andreon, 
S., Davoust, E., \& Poulain, P.\ 1997, \aaps, 126, 67 

\bibitem[Andreon et al.(1996)]{} Andreon, S., Davoust, 
E., Michard, R., Nieto, J.-L., \& Poulain, P.\ 1996, \aaps, 116, 429 

\bibitem[Barazza, Binggeli, \& Jerjen(2003)]{} Barazza, 
F.~D., Binggeli, B., \& Jerjen, H.\ 2003, \aap, 407, 121

\bibitem[Barazza, Binggeli, \& Jerjen(2002)]{} Barazza, 
F.~D., Binggeli, B., \& Jerjen, H.\ 2002, \aap, 391, 823

\bibitem[Beers, Flynn, \& Gebhardt(1990)]{} Beers, 
T.~C., Flynn, K., \& Gebhardt, K.\ 1990, \aj, 100, 32 

\bibitem[Bekki, Couch, \& Shioya(2002)]{} Bekki, K., 
Couch, W.~J., \& Shioya, Y.\ 2002, \apj, 577, 651

\bibitem[Bekki(2001)]{} Bekki, K.\ 2001, \apj, 546, 189 

\bibitem[]{} Bertin, E. \& Arnouts, S. \ 1996, A\&AS, 117, 393 

\bibitem[Binggeli \& Popescu(1995)]{} Binggeli, B.~\& 
Popescu, C.~C.\ 1995, \aap, 298, 63

\bibitem[Binggeli \& Cameron(1991)]{} Binggeli, B.~\& 
Cameron, L.~M.\ 1991, \aap, 252, 27 

\bibitem[Binggeli et al.(1988)]{} Binggeli, B., Sandage, 
A., \& Tammann, G.~A.\ 1988, \araa, 26, 509

\bibitem[Binggeli, Sandage, \& Tammann(1985)]{} 
Binggeli, B., Sandage, A., \& Tammann, G.~A.\ 1985, \aj, 90, 1681

\bibitem[Biviano \& Katgert(2004)]{} Biviano, A.~\& 
Katgert, P.\ 2004, \aap, 424, 779 

\bibitem[Caldwell \& Bothun(1987)]{} Caldwell, N.~\& 
Bothun, G.~D.\ 1987, \aj, 94, 1126

\bibitem[Caon, Capaccioli, \& D'Onofrio(1993)]{} Caon, 
N., Capaccioli, M., \& D'Onofrio, M.\ 1993, \mnras, 265, 1013

\bibitem[de Blok et al.(1995)]{} de Blok, W.~J.~G., van 
der Hulst, J.~M., \& Bothun, G.~D.\ 1995, \mnras, 274, 235

\bibitem[Dekel \& Silk(1986)]{} Dekel, A.~\& Silk, J.\ 
1986, \apj, 303, 39

\bibitem[De Young \& Gallagher(1990)]{} De Young, 
D.~S.~\& Gallagher, J.~S.\ 1990, \apjl, 356, L15 

\bibitem[Driver et al.(1994)]{} Driver, S.~P., 
Phillipps, S., Davies, J.~I., Morgan, I., \& Disney, M.~J.\ 1994, \mnras, 
266, 155

\bibitem[Durrell(1997)]{} Durrell, P.~R.\ 1997, \aj, 
113, 531

\bibitem[Freeman(1970)]{} Freeman, K.~C.\ 1970, \apj, 
160, 811

\bibitem[Geha, Guhathakurta, \& van der Marel(2003)]{} 
Geha, M., Guhathakurta, P., \& van der Marel, R.~P.\ 2003, \aj, 126,
1794

\bibitem[Godwin, Metcalfe, \& Peach(1983)]{} Godwin, 
J.~G., Metcalfe, N., \& Peach, J.~V.\ 1983, \mnras, 202, 113 

\bibitem[Graham \& Guzm{\' a}n(2003)]{} Graham, A.~W.~\& 
Guzm{\' a}n, R.\ 2003, \aj, 125, 2936

\bibitem[Gunn \& Gott(1972)]{} Gunn, J.~E.~\& Gott, 
J.~R.~I.\ 1972, \apj, 176, 1

\bibitem[Guti{\' e}rrez et al.(2004)]{} Guti{\' e}rrez, 
C.~M., Trujillo, I., Aguerri, J.~A.~L., Graham, A.~W., \& Caon, N.\ 2004, 
\apj, 602, 664 



\bibitem[Iglesias-P{\' a}ramo et al.(2003)]{} 
Iglesias-P{\' a}ramo, J., Boselli, A., Gavazzi, G., Cortese, L., \& 
V{\'{\i}}lchez, J.~M.\ 2003, \aap, 397, 421 

\bibitem[Iglesias-P{\' a}ramo et al.(2002)]{} 
Iglesias-P{\' a}ramo, J., Boselli, A., Cortese, L., V{\'{\i}}lchez, J.~M., 
\& Gavazzi, G.\ 2002, \aap, 384, 383

\bibitem[Jerjen, Kalnajs, \& Binggeli(2000)]{} Jerjen, 
H., Kalnajs, A., \& Binggeli, B.\ 2000, \aap, 358, 845 

\bibitem[Jerjen \& Binggeli(1997)]{} Jerjen, H.~\& 
Binggeli, B.\ 1997, ASP Conf.~Ser.~116: The Nature of Elliptical Galaxies; 
2nd Stromlo Symposium, 239

\bibitem[Kormendy(1985)]{} Kormendy, J.\ 1985, \apj, 295, 73 
 
\bibitem[Lin \& Faber(1983)]{} Lin, D.~N.~C.~\& Faber, 
S.~M.\ 1983, \apjl, 266, L21 

\bibitem[]{} Mastropietro, C., Moore, B., Mayer, L., Debattista,
  V. P., Piffaretti, R., \& Stadel, J. \ 2004, astro-ph/0411648 

\bibitem[Mayer et al.(2001a)]{} Mayer, L., Governato, F., 
Colpi, M., Moore, B., Quinn, T., Wadsley, J., Stadel, J., \& Lake, G.\ 
2001a, \apj, 559, 754 

\bibitem[Mayer et al.(2001b)]{} Mayer, L., Governato, F., 
Colpi, M., Moore, B., Quinn, T., Wadsley, J., Stadel, J., \& Lake, G.\ 
2001b, \apjl, 547, L123

\bibitem[Mobasher et al.(2003)]{} Mobasher, B., et al.\ 
2003, \apj, 587, 605 

\bibitem[Moore et al.(2000)]{} Moore, B., Gelato, S., 
Jenkins, A., Pearce, F.~R., \& Quilis, V.\ 2000, \apjl, 535, L21 

\bibitem[Moore, Lake, Quinn, \& Stadel(1999)]{} Moore, 
B., Lake, G., Quinn, T., \& Stadel, J.\ 1999, \mnras, 304, 465 

\bibitem[Moore, Lake, \& Katz(1998)]{} Moore, B., Lake, 
G., \& Katz, N.\ 1998, \apj, 495, 139 

\bibitem[Moore et al.(1996)]{} Moore, B., Katz, N., 
Lake, G., Dressler, A., \& Oemler, A.\ 1996, \nat, 379, 613

\bibitem[Navarro, Frenk, \& White(1997)]{} Navarro, 
J.~F., Frenk, C.~S., \& White, S.~D.~M.\ 1997, \apj, 490, 493

\bibitem[Patterson \& Thuan(1996)]{} Patterson, R.~J.~\& 
Thuan, T.~X.\ 1996, \apjs, 107, 103

\bibitem[Pedraz et al.(2002)]{} Pedraz, S., Gorgas, J., 
Cardiel, N., S{\' a}nchez-Bl{\' a}zquez, P., \& Guzm{\' a}n, R.\ 2002, 
\mnras, 332, L59

\bibitem[Phillipps et al.(1998)]{} Phillipps, S., 
Driver, S.~P., Couch, W.~J., \& Smith, R.~M.\ 1998, \apjl, 498, L119 

\bibitem[Poggianti et al.(2001)]{} Poggianti, B.~M., et 
al.\ 2001, \apj, 562, 689

\bibitem[Pracy et al.(2004)]{} Pracy, M.~B., De Propris, 
R., Driver, S.~P., Couch, W.~J., \& Nulsen, P.~E.~J.\ 2004, \mnras, 352, 
1135

\bibitem[Quilis, Moore, \& Bower(2000)]{} Quilis, V., 
Moore, B., \& Bower, R.\ 2000, Science, 288, 1617

\bibitem[Richer \& McCall(1995)]{} Richer, M.~G.~\& 
McCall, M.~L.\ 1995, \apj, 445, 642

\bibitem[Ryden et al.(1999)]{} Ryden, B.~S., Terndrup, 
D.~M., Pogge, R.~W., \& Lauer, T.~R.\ 1999, \apj, 517, 650

\bibitem[]{} S\'anchez-Janssen, R., Iglesias-P\'aramo, J.,
  Mu\~noz-Tu\~n\'on, C., Aguerri, J. A. L. ~\& V\'{\i}lchez, J. M. \
  2005, \aap, in press

\bibitem[Sandage \& Binggeli(1984)]{} Sandage, A.~\& 
Binggeli, B.\ 1984, \aj, 89, 919 

\bibitem[Secker, Harris, \& Plummer(1997)]{} Secker, J., 
Harris, W.~E., \& Plummer, J.~D.\ 1997, \pasp, 109, 1377

\bibitem[Sersic(1968)]{} Sersic, J.~L.\ 1968, Atlas de 
Galaxes Australes; Vol.~Book; Page 1, 0 

\bibitem[Silich \& Tenorio-Tagle(2001)]{} Silich, S.~\& 
Tenorio-Tagle, G.\ 2001, \apj, 552, 91

\bibitem[Skillman, Kennicutt, \& Hodge(1989)]{} 
Skillman, E.~D., Kennicutt, R.~C., \& Hodge, P.~W.\ 1989, \apj, 347,
875

\bibitem[Thompson \& Gregory(1993)]{} Thompson, L.~A.~\& 
Gregory, S.~A.\ 1993, \aj, 106, 2197

\bibitem[Toomre \& Toomre(1972)]{} Toomre, A.~\& Toomre, 
J.\ 1972, \apj, 178, 623 

\bibitem[Trentham(1998)]{} Trentham, N.\ 1998, \mnras, 
293, 71

\bibitem[Trujillo \& Aguerri(2004)]{} Trujillo, I.~\& 
Aguerri, J.~A.~L.\ 2004, \mnras, 355, 82

\bibitem[Trujillo et al.(2002)]{} Trujillo, I., Aguerri, 
J.~A.~L., Guti{\' e}rrez, C.~M., Caon, N., \& Cepa, J.\ 2002, \apjl, 573, 
L9 

\bibitem[Trujillo, Aguerri, Guti{\' e}rrez, \& 
Cepa(2001)]{} Trujillo, I., Aguerri, J.~A.~L., Guti{\' 
e}rrez, C.~M., \& Cepa, J.\ 2001a, \aj, 122, 38

\bibitem[Trujillo, Aguerri, Cepa, \& Guti{\' 
e}rrez(2001)]{} Trujillo, I., Aguerri, J.~A.~L., Cepa, 
J., \& Guti{\' e}rrez, C.~M.\ 2001b, \mnras, 328, 977 

\bibitem[Trujillo, Aguerri, Cepa, \& Guti{\' 
e}rrez(2001)]{} Trujillo, I., Aguerri, J.~A.~L., Cepa, 
J., \& Guti{\' e}rrez, C.~M.\ 2001c, \mnras, 321, 269

\bibitem[van Zee, Skillman, \& Haynes(2004)]{} van Zee, 
L., Skillman, E.~D., \& Haynes, M.~P.\ 2004a, \aj, 128, 121 

\bibitem[van Zee, Skillman, ]{} van Zee, 
L., Barton, E. J., \& Skillman, E.~D. \ 2004b, \aj, in press 

\bibitem[van Zee, Salzer, \& Skillman(2001)]{} van Zee, 
L., Salzer, J.~J., \& Skillman, E.~D.\ 2001, \aj, 122, 121

\bibitem[van Zee(2000)]{} van Zee, L.\ 2000, \aj, 119, 
2757
 
\end{thebibliography}
\end{document}